\documentclass[12pt,nofootinbib]{revtex4}%
\usepackage{epsfig,graphicx,amsmath}
\usepackage{amstext,amsbsy,amscd,amsopn}
\usepackage{amsmath}
\usepackage{amsfonts}
\usepackage{amssymb}
\usepackage{graphicx}%
\begin{document}
\title{Effects of intersegmental transfers on target location by proteins}
\author{Michael Sheinman and Yariv Kafri}
\affiliation{Department of Physics, Technion-Israel Institute of Technology, 32000 Haifa, Israel}

\begin{abstract}
We study a model for a protein searching for a target, using facilitated
diffusion, on a DNA molecule confined in a finite volume. The model includes
three distinct pathways for facilitated diffusion: (a) \textit{sliding} - in
which the protein diffuses along the contour of the DNA (b) \textit{jumping} -
where the protein travels between two sites along the DNA by three-dimensional
diffusion, and finally (c) \textit{intersegmental transfer} - which allows the
protein to move from one site to another by transiently binding both at the
same time. The typical search time is calculated using scaling arguments which
are verified numerically. Our results suggest that the inclusion of
intersegmental transfer (i) decreases the search time considerably (ii) makes
the search time much more robust to variations in the parameters of the model
and (iii) that the optimal search time occurs in a regime very different than
that found for models which ignore intersegmental transfers. The behavior we
find is rich and shows surprising dependencies, for example, on the DNA length.

\end{abstract}
\date{\today}
\maketitle

\section{Introduction}

Many biological processes depend on the ability of proteins to locate specific
DNA sequences on time scales ranging from seconds to minutes. Examples include
gene expression and repression, DNA replication and others~\cite{ABLRRW94}.
Naively, one might expect the protein to search for its target using only
three-dimensional diffusion\footnote{In this paper we only consider proteins
whose motion is diffusive and not directed (directed motion could result from
consumption of, for example, chemical energy and is discussed in
\cite{LBMV2008}).}. Neglecting interactions of the protein with the
environment and the DNA (apart from the target site) one then finds, using
results first obtained by Smoluchowski \cite{S17}, that the average search
time, $t^{search}$, is given by:
\begin{equation}
t^{search}\sim\frac{\Lambda^{3}}{D_{3}r}\;. \label{t3D}%
\end{equation}
Here $D_{3}$ is the three-dimensional diffusion constant of the protein, $r$
is the target size and $\Lambda^{3}$ is the volume that needs to be searched.
Assuming a target size of the order of a base-pair $r\approx0.34nm$, a typical
nucleus (or bacteria) size of $\Lambda\sim10^{3}nm$ and using the measured
three-dimensional diffusion coefficient for a GFP protein \textit{in vivo,}
$D_{3}\sim10^{7}{nm}^{2}/s$ \cite{ESWSL99}, one finds $t^{search}$ of the
order of hundreds of seconds. If $N$ proteins are searching for the same
target the search time is given by\footnote{The relation between the search
time $t^{search}$ for one protein and search time $t_{N}^{search}$ for $N$
proteins remains unchanged throughout the paper.} $t_{N}^{search}\simeq
t^{search}/N$. This suggests that about $10$ proteins could find a target in
reasonable times for cells to function properly.

In real systems, due to the interactions of proteins with non-specific DNA
sequences and the environment \cite{LR72}, the picture is more complex.
Indeed, \textit{in vitro} experiments have suggested that mechanisms other
than three-dimensional diffusion are used by many proteins to locate their
targets \cite{RSB70,RBC70}. These strategies have been studied and debated
extensively both in the context of \textit{in vivo }%
\cite{BWH81,SM2004,HGS2006,HS2007} and \textit{in vitro} systems
\cite{BB76,BWH81,HM2004,BZ2004,Z2005,LAM2005,HGS2006} and are believed, in
general, to allow for search times which are faster than that given by Eq.
(\ref{t3D}).

\begin{figure}[ptb]
\begin{center}
\includegraphics[
height=3.218in, width=4.9286in ]{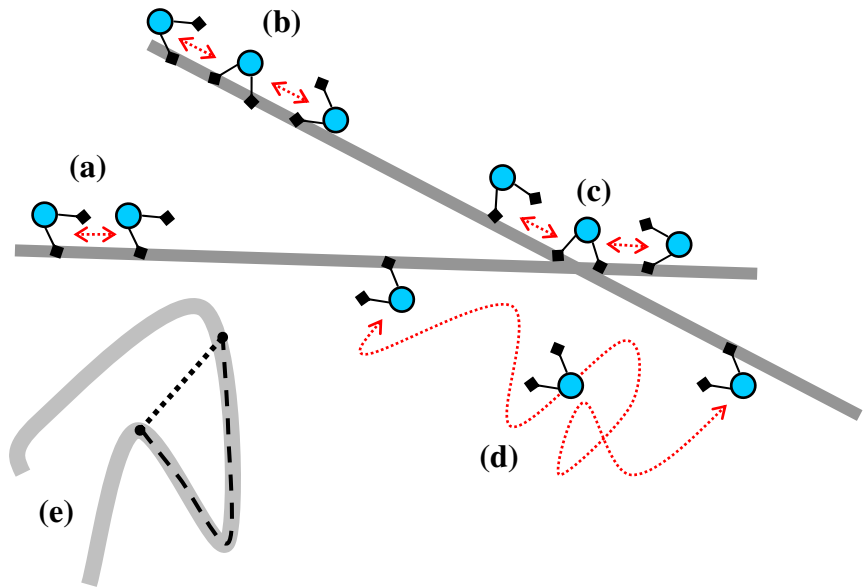}
\end{center}
\caption{Schematic plots illustrating the different mechanisms that can
participate in the facilitated diffusion process. Here dashed arrows represent
different protein moves, the solid curve represents the DNA and a small circle
with two legs indicates a protein with two binding domains. The figure shows
(a) sliding, (b) a correlated intersegmental transfer, (c) an uncorrelated
intersegmental transfer, (d) jumping. The distinction between (b) and (c) is
defined in Sec. \ref{Pure intersegmental transfer}. (e) The dashed (dotted)
line represents a one-dimensional (three-dimensional) distance.}%
\label{fig1}%
\end{figure}

Historically, the first strategy that was proposed combines one-dimensional
diffusion (sliding) over the DNA with intervals of three-dimensional diffusion
(typically called jumping in this context) \cite{AD68,BWH81} (see Fig.
\ref{fig1}). Each individual search mechanism, when applied alone, has
shortcoming and advantages over the other. When using only three-dimensional
diffusion, the number of \textit{new} three dimensional positions probed grows
linearly in time but the protein spends much time probing sites where there is
no DNA present. In contrast, during one-dimensional diffusion the protein is
constantly bound to the DNA but suffers from a slow increase in the number of
\textit{new} positions probed as a function of time ($\sim t^{1/2}$, where $t$
denotes time) \cite{H95Volume1}. As shown, for example, in Refs.
\cite{AD68,BWH81} by intertwining one and three dimensional search strategies
and \textit{tuning the properties of both} one can in fact decrease the search
time significantly\footnote{Clearly, a pure one-dimensional search strategy is
not efficient due to the slow diffusive search along the DNA, $t^{search}%
\sim\frac{L^{2}}{D_{1}}\sim O\left(  hours\right)  $, where $L\sim10^{6}nm$ is
the genome length and $D_{1}$ is the one-dimensional diffusion coefficient
that was measured indirectly \cite{BB76} and directly \cite{WAC2006,ELX2007}
to be much smaller than three-dimensional diffusion coefficient $D_{3}%
\sim10^{7}\frac{nm^{2}}{s}$ \cite{ESWSL99}.}.

The combined strategy, while better than the pure search strategies, comes at
a cost of being sensitive to changes in the properties of either the
three-dimensional or one-dimensional diffusive processes. For example, as we
argue below, the typical search time changes exponentially in the square root
of the ionic strength. Moreover, given the many constraints on the protein to
function it is very restrictive to demand optimization for the search process.
Indeed, equilibrium measurements \cite{K-HRBOCNVH77} and recent single
molecule experiment \cite{WAC2006,ELX2007} on the Lac repressor protein
suggest that the search process may not be in general optimized for this
search strategy.

A third mechanism which was suggested to speed the search time is
intersegmental transfer (IT) \cite{HRGW75,BC75}. During an IT the protein
moves from one site to another by transiently binding both at the same time.
In principle the new site can be either close along the one-dimensional DNA
sequence (or chemical distance) or distant (see Fig. \ref{fig2}). This
mechanism is likely to be relevant for the proteins that have more than one
binding domain like the Lac repressor \cite{FM92,RC92}, GRdbd \cite{LN97} and
SfiI enzyme \cite{EWWH99}. However, it could also occur in proteins with a
single binding site in locations where the DNA crosses itself . To date we are
aware of direct evidence for IT only for RNA polymerase \cite{BGZY99}.
However, measurement of the dissociation rate from a labeled (operator) DNA
site of the rat glucocorticoid receptor \cite{LN97}, CAP and Lac repressor
\cite{FC84} revealed significant dependence on the DNA concentration in the
solvent, a possible explanation for which is IT. Some theoretical work has suggested that \textit{in vivo}, when
the DNA concentration is much larger than \textit{in vitro} experiments, IT
may play a determinative role \cite{BWH81,LAM2005,HS2007}. These studies focus
on the ITs resulting from the DNA dynamics and consider the protein to be
point like.

In this paper we present a rather comprehensive study of the effects of ITs on
the search process for a DNA molecule confined in a finite volume, similar to
the \textit{in vivo} scenario. Our work complements previous ones by
explicitly accounting for the size of the protein and considering two limiting
cases: (i) DNA which is completely static during the search process and (ii)
DNA whose motion is quicker than that of the protein's motion along the DNA.
Using scaling arguments backed by numerics we obtain expressions for
$t^{search}$, and the optimal search time (obtained by tuning parameters such
as the DNA-protein affinity). A central conclusion of this paper is that the
search time is much more robust to variations in parameters when ITs are
allowed\footnote{Of course, this fact may be both advantageous and
disadvantageous for the cell. In some cases the cell needs transcription
factors whose kinetic (and, therefore equilibrium) properties do depend on the
environment and in other cases it doesn't.}. This is to the extent that in
some cases any finite jumping rate can have a negative influence on the search
time. In particular, the optimal search time is found to occur for parameter
regimes very different than the canonical one (see Sec.
\ref{Sliding and jumping}) found in models which ignore ITs. Perhaps most
important, as we show, our work suggests that ITs could explain recent
findings which indicate a much higher affinity of the TF Lac repressor to the
DNA than required by an optimal search strategy which uses only sliding and
jumping \cite{WAC2006,ELX2007,K-HRBOCNVH77}.

The scaling dependence of the search time on different parameters is
rich and very different from regular facilitated diffusion
(involving only sliding and jumping). Consider, for example, the
dependence of the search process on the length of the DNA, $L$ for a
DNA confined in a volume $\Lambda^{3}$. Using only sliding and
jumping the regime typically thought to be relevant to experiment
has a linear dependence of the search time on the DNA\ length $L$. A
Smoluchowski-like search time is independent of $L$. In contrast,
when ITs are allowed we find different behavior. We estimate that
the regime most relevant to \textit{in vivo} experiments (in
prokaryotic organisms) occurs when the dependence on the length of
the DNA is \textit{weak}. For example, when a searches are performed
using only ITs the search time can be independent of $L$ or scales
as $\sqrt{L}$ depending on the DNA's dynamics. The scaling behaviors
relics on the confinement of the DNA in a finite volume (shown in
detail in Fig. \ref{Scheme}) and could be used as experimental
probes for the existence of ITs.

The paper is organized as follows: Sec. \ref{Sliding and jumping} briefly
reviews the main arguments used to analyze searches that combines only sliding
and jumping. In Sec. \ref{Pure intersegmental transfer}\ the average search
time is calculated for the case of a strategy based only on ITs for both
quenched and annealed DNA. In Sec. \ref{Intersegmental transfer and sliding} a
search process that includes ITs and sliding is considered. Sec.
\ref{Intersegmental transfer and jumping} considers the possibility that the
protein can unbind from the DNA (jump) and perform ITs. Sec.
\ref{Intersegmental transfer, sliding and jumping} studies a model with all
three mechanisms. Finally, in Sec. \ref{Application to the Lac repressor} we
discuss possible scenarios for the Lac repressor and summarize in Sec.
\ref{Summary}.

\section{Sliding and jumping\label{Sliding and jumping}}

To set the stage for a discussion of the effects of IT we consider a search
process which uses only sliding and jumping. The discussion follows Refs.
\cite{SM2004} and \cite{HM2004} closely. We imagine a single protein searching
for a single target located on the DNA. The search is composed of a series of
intervals of one-dimensional diffusion along the DNA (sliding) and
three-dimensional diffusion in the solution (jumping). The typical time of
each is denoted by $\tau_{1}$ and $\tau_{3}$ respectively. Following a jump,
the protein is assumed to associate on a new randomly chosen location along
the DNA. While this approach is somewhat simplistic for jumps occurring in
two-dimensions and below, for three dimensions, which case we consider, it is
well suited \cite{CBTVK2007}.

\begin{figure}[ptb]
\begin{center}
\includegraphics[
height=2.6178in,
width=4.0871in
]{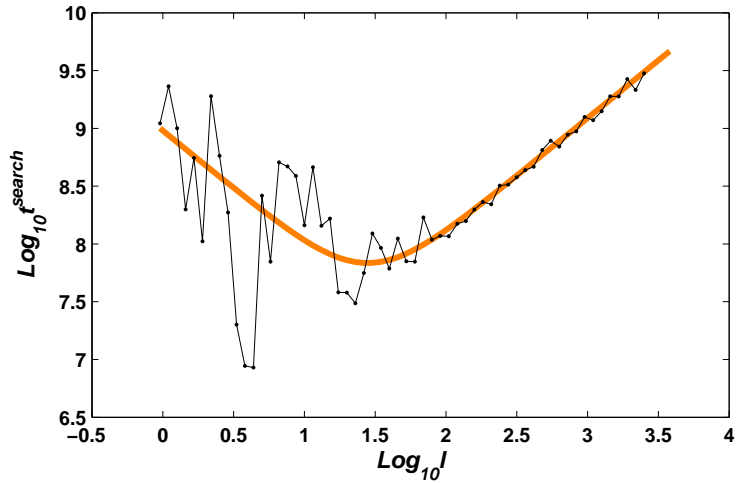}
\end{center}
\caption{The search time $t^{search}$ is shown as a function of the antenna
length, $l$. The thin line represents the results from numerical simulations
while the bold one is given by Eq. (\ref{ts1}). Numerics were performed on a
DNA embedded in a finite volume with a frozen configuration. The length of the
DNA was taken to be $1224000$ lattice constants and $D_{3}=D_{1}=1$ (see
details in Appendix \ref{Appendix B}). Similar results were obtained for
different values of $D_{1}$ and $D_{3}$.}%
\label{SMfig}%
\end{figure}

Under these assumptions, during each sliding event the protein covers a
typical length $l$, where $l\sim\sqrt{D_{1}\tau_{1}}$ (often called
the\textit{\ antenna }size) \cite{H95Volume1}. Since correlations between the
locations of the protein before and after the jump are neglected, the search
process, completed when roughly all the DNA is scanned, is separated into
\begin{equation}
N_{r}\sim\frac{L}{l_{s}} \label{NrGen}%
\end{equation}
rounds of sliding and jumping. Here $l_{s}$ is the typical length scanned by
the protein during a round. If during the slide the protein does not skip
sites on the DNA $l_{s}\sim l$ (the distinction between $l_{s}$ and $l$ will
become apparent when ITs are introduced). The total time needed to find a
specific site is then: \ \
\begin{equation}
t^{search}=N_{r}\tau_{r}\;, \label{SM}%
\end{equation}
with $\tau_{r}=\tau_{1}+\tau_{3}$. Using Eqs. (\ref{NrGen}) and (\ref{SM}) one
obtains%
\begin{equation}
t^{search}\sim\frac{L}{l_{s}}(\tau_{1}+\tau_{3})\sim\frac{L}{\sqrt{2D_{1}}%
}\left(  \sqrt{\tau_{1}}+\frac{\tau_{3}}{\sqrt{\tau_{1}}}\right)  \;.
\label{ts1}%
\end{equation}
Furthermore, it is easy to argue (see Appendix \ref{Appendix A}) that
\begin{equation}
\tau_{3}\sim\frac{\Lambda^{3}}{D_{3}L}\;. \label{t3}%
\end{equation}
In Fig. \ref{SMfig} a comparison between the presented scaling arguments and a
numerical simulation of a search that explicitly includes sliding on a DNA
with a frozen configuration in a finite volume and three dimensional diffusion
is shown (see Appendix \ref{Appendix B} for details of the numerics). The
excellent agreement justifies many of the simplifications made, in particular,
the neglect correlation between the initial and final location of the jump.
Throughout the \ paper we assume this always holds (see Appendix
\ref{Appendix B}). 

The analysis leads to a richer range of possible behaviors than found in Eq.
(\ref{t3D}), where the search time depends only on the volume in which the DNA
is embedded \cite{HGS2006}. Here, in contrast, three regimes are found: (i)
For $\tau_{1}\ll\tau_{3}$ there is no dependence on $L$ and the search time is
given to a good approximation by Eq. (\ref{t3D}). (ii) For $\frac{L^{2}}%
{D_{1}}\gg\tau_{1}\gg\tau_{3}$ the dependence on the DNA\ length is linear.
This is the regime typically considered relevant for experiments. (iii) For
$\frac{L^{2}}{D_{1}}\ll\tau_{1}$ one finds $t^{search}\propto L^{2}$.

It is natural to ask which $\tau_{1}$ optimizes $t^{search}$. Using Eq.
(\ref{ts1}) it is easy to verify that
\begin{equation}
\left(  \tau_{1}^{opt}\right)  _{0}=\tau_{3}\;, \label{t1opt0}%
\end{equation}
where $0$ denotes a value obtained with no ITs. Alternatively, one can
consider an optimal antenna size $(l_{opt})_{0}=\sqrt{2D_{1}\tau_{3}}$. When
this condition is met, the total search time scales as
\begin{equation}
t_{opt}^{search}=\sqrt{\frac{\tau_{3}}{D_{1}}}L\sim\sqrt{\frac{\Lambda^{3}%
L}{D_{1}D_{3}}}\;. \label{tSM}%
\end{equation}
Note that the $\sqrt{L}$ dependence is obtained by optimizing, say $\tau_{1}$,
as $L$ is varied.

This model, \textit{at the optimal} $\tau_{1}$ and assuming known values for
$D_{1}$, $L$ and $\tau_{3}$, predicts reasonable search times \textit{in vivo
}and is commonly assumed to give a possible explanation for the two order of
magnitude\ difference between the experiments \textit{in vitro} and Eq.
(\ref{t3D}).

Within the model the optimal search process requires fine tuning of the
antenna size, $l$, as a function of the parameters $D_{1}$ and $\tau_{3}$.
These parameters depend on various cell and environmental conditions such as
the size of the cell, the DNA length, the ionic strength etc. The dependence
can be quite significant: for example, the parameter $\frac{\tau_{3}}{\tau
_{1}}$ has an exponential dependence on the square root of the ionic strength
\cite{LR75}. Deviations of this parameter from the optimum value might be
crucial to the search time since $\frac{t^{search}}{t_{opt}^{search}}=\frac
{1}{2}\left(  \sqrt{\frac{\tau_{3}}{\tau_{1}}}+\sqrt{\frac{\tau_{1}}{\tau_{3}%
}}\right)  $. Indeed, a strong dependence of the search time on the ionic
strength was found in \textit{in vitro} experiments \cite{RBC70}.
Interestingly, \textit{in vivo}, when the DNA is densely packed, no effect of
the ionic strength on the efficiency of the Lac repressor was revealed
\cite{RCMKAFR87}. Other experiments also suggest that $\tau_{1}$ is not
optimized. In particular, equilibrium measurements \cite{K-HRBOCNVH77}, as
well as recent single molecule experiment \cite{WAC2006,ELX2007}, find a value
of $\tau_{1}$\ for \textit{dimeric} Lac repressor that is much larger than the
predicted optimum $\tau_{3}$ \textit{in vivo}.

The lack of sensitivity to the ionic strength \textit{in vivo} and the rapid
search times found for the Lac repressor, even with very large values of
$\tau_{1}$, suggest that other processes, apart from jumping and sliding, are
involved in the search process. These seem to be more important \textit{in
vivo} than \textit{in vitro}. In the next section we show that a search
process which uses ITs modifies the behavior found for searches which use only
sliding and jumping in a significant manner. In particular the problems
encountered above (e.g., high sensitivity to the antenna length, very long and
non-optimal measured antennas etc.), are largely eliminated when ITs are included.

\section{Pure intersegmental transfer \label{Pure intersegmental transfer}}

Before turning to the full problem of a search which uses sliding, jumping and
ITs we will consider a series of simplified models. Within the\ first model,
considered in this section, the protein can only perform ITs. We will see that
already at this level many of the problems of the search discussed above,
which uses only sliding and jumping, are resolved to a large extent.

To model ITs we consider a protein with two binding sites. The protein can
either have one site bound to the DNA or perform an IT to a new location by
having both binding sites bound to the DNA (see Fig. \ref{fig1}). The DNA is
scanned for the target by the binding sites, each checking a length $b$ when
bound (note that since the protein has to align with the DNA sequence, $b$ is
of the order of a length of a single base-pair). A possible motivation for
this picture is, for example, the tetrameric structure of the Lac repressor.
However, as will be evident many results also apply to proteins with different shapes.

Motivated by DNA in cells, we consider a DNA molecule which is densely packed
in a small volume. In typical systems the DNA has a total length of
$L\sim10^{6}nm$, a persistence length $L_{0}\sim50nm$, a cross section radius
$\rho\sim1nm$ and is contained in a volume of $\Lambda^{3}\sim10^{9}nm^{3}$.
The typical distance between segments of DNA of length $L_{0}$ is therefore
much smaller than $L_{0}$: $\frac{\Lambda^{3}}{L/L_{0}}\ll L_{0}^{3}$. Under
these conditions, using $\Lambda\gg L_{0}$, it is easy to check that the
radius of gyration of free DNA, which is of the order of $L_{0}\sqrt{\frac
{L}{L_{0}}}$ is much larger than the cell size $\Lambda$ - the DNA is densely
packed even though its fractional volume, $L\rho^{2}/\Lambda^{3}$, in the
container is small (about one percent). By way of comparison, typical protein
sizes are in the range $R\mathbf{\sim}1-10nm$, much smaller than the DNA's
persistence length. Although \textit{in vivo} the packing has a more
complicate structure than we consider, we expect similar behavior to occur
also there.

As stated above the protein moves by first being bound with only one binding
site and then with both. The typical time for this, defined by $\delta
=\tau_{b}+\tau_{IT}$, is the sum of the typical time that protein probes a
length $b$ (by being bound with one domain) and the time that the protein is
bound with both binding domains to the DNA while performing an IT\footnote{We
take $\delta$ independent of parameters such as cell size $\Lambda$ and the
DNA length, $L$. This is justified in a regime where most ITs are close along
the chemical coordinate of the DNA.}. We assume that the protein moves (for
example, using both legs of the Lac repressor) to a random position located at
a distance smaller or equal to $R$, the size of the protein, from it (see Fig.
\ref{fig1})\footnote{Different scenarios are considered at the end of the
section.}. Defining a ``chemical'' coordinate $x$ which runs along the length
of the DNA the protein can either perform moves from its location $x$ to the
interval $[x-R,x+R]$ (we refer to these as ``correlated ITs'' (CITs)) or reach
distant sites along the chemical coordinate available through the structure of
the packed DNA.

Under the above conditions it is easy to verify (see Appendix \ref{Appendix C}%
) that almost all ITs performed by the protein are either correlated moves or
performed to a coordinate along the DNA whose distance from its previous
location is bigger than $\frac{\Lambda^{2}}{L_{0}}$ (but smaller than $L$). We
call these steps \textquotedblleft uncorrelated ITs\textquotedblright\ (UITs)
(see Fig. \ref{fig1}(c)). In other words, one can safely neglect the
possibility that the protein will move using ITs to a chemical distance larger
than $R$ and smaller than $\frac{\Lambda^{2}}{L_{0}}$.

Our main interest is the typical search time. For this purpose it is useful to
define $\lambda$ - the average length that the protein travels before
performing an UIT. On chemical distances larger than $R$ but smaller than
$\lambda$ the motion is effectively diffusive in one dimension with a
diffusion coefficient $D_{1eff}$ $\sim\frac{R^{2}}{\delta}$. On chemical
distance scales larger than $\lambda$ and smaller than $L$ the motion is
controlled by UITs. Due to the three-dimensional nature of each UIT one
expects correlations between different UITs to be negligible. We verify this
assumption later using numerical simulations.

From the discussion and using a language similar to that of Sec. II the search
process can be described as a sequence of
\begin{equation}
N_{r}\sim\frac{L}{l_{s}} \label{Nr}%
\end{equation}
\textit{rounds} of correlated ITs where $l_{s}$ is the length \textit{scanned}
by the protein during each round (namely between two subsequent UITs). The
typical time of each round is%
\begin{equation}
\tau_{r}\sim\frac{\lambda^{2}}{D_{1eff}}\sim\left(  \frac{\lambda}{R}\right)
^{2}\delta\;. \label{Tr}%
\end{equation}

In general while performing CITs the protein can miss regions of the DNA by
skipping over them. Since each segment of size $R$ is visited $\sqrt
{\frac{\tau_{r}}{\delta}}\sim\frac{\lambda}{R}$ times \cite{H95Volume1}, when
$\frac{\lambda}{R}\gg\frac{R}{b}$ the walk is recurrent and no sites are
skipped so that $l_{s}\sim\lambda$. In contrast, when $\frac{\lambda}{R}%
\ll\frac{R}{b}$ the walk is not recurrent and $l_{s}\sim\frac{\lambda}%
{R/b}\frac{\lambda}{R}\sim\frac{\lambda^{2}}{R^{2}}b$. Therefore the
recurrence length,
\begin{equation}
l_{R}\sim\frac{R^{2}}{b}\;, \label{lR}%
\end{equation}
separates between two regimes
\begin{equation}
l_{s}\sim\left\{
\begin{array}
[c]{c}%
\frac{\lambda^{2}}{l_{R}}\hspace{0.5in}\lambda\ll l_{R}\\
\lambda\hspace{0.6in}\lambda\gg l_{R}%
\end{array}
\right.  \;, \label{2Gen}%
\end{equation}
the first transient and the second recurrent.

Using Eqs. (\ref{SM}), (\ref{Nr}), (\ref{Tr}) and (\ref{2Gen}) the typical
search time is obtained%
\begin{equation}
t^{search}\sim\frac{L}{l_{s}}\left(  \frac{\lambda}{R}\right)  ^{2}\delta
\sim\left\{
\begin{array}
[c]{c}%
\frac{L}{b}\delta\;\;\;\;\;\;\;\lambda\ll l_{R}\\
\frac{L\lambda}{R^{2}}\delta\;\;\;\;\;\;\lambda\gg l_{R}%
\end{array}
\right.  \;. \label{1Gen}%
\end{equation}
To complete the expression one needs to evaluate $\lambda$. Its value depends
on various parameters and, in particular, the time scale which characterize
the motion of the DNA. As discussed in the introduction we consider two
extreme regimes - quenched DNA and annealed DNA. In both cases $\lambda$ can
be evaluated from an intermediate quantity, $p$, the probability that the
protein can make an UIT from a specific location $x$ on the DNA. Since this
quantity is independent of the DNA's motion we estimate it first before
turning to the two regimes.

\begin{figure}[ptb]
\begin{center}
\includegraphics[
height=2.1897in,
width=6.1799in
]{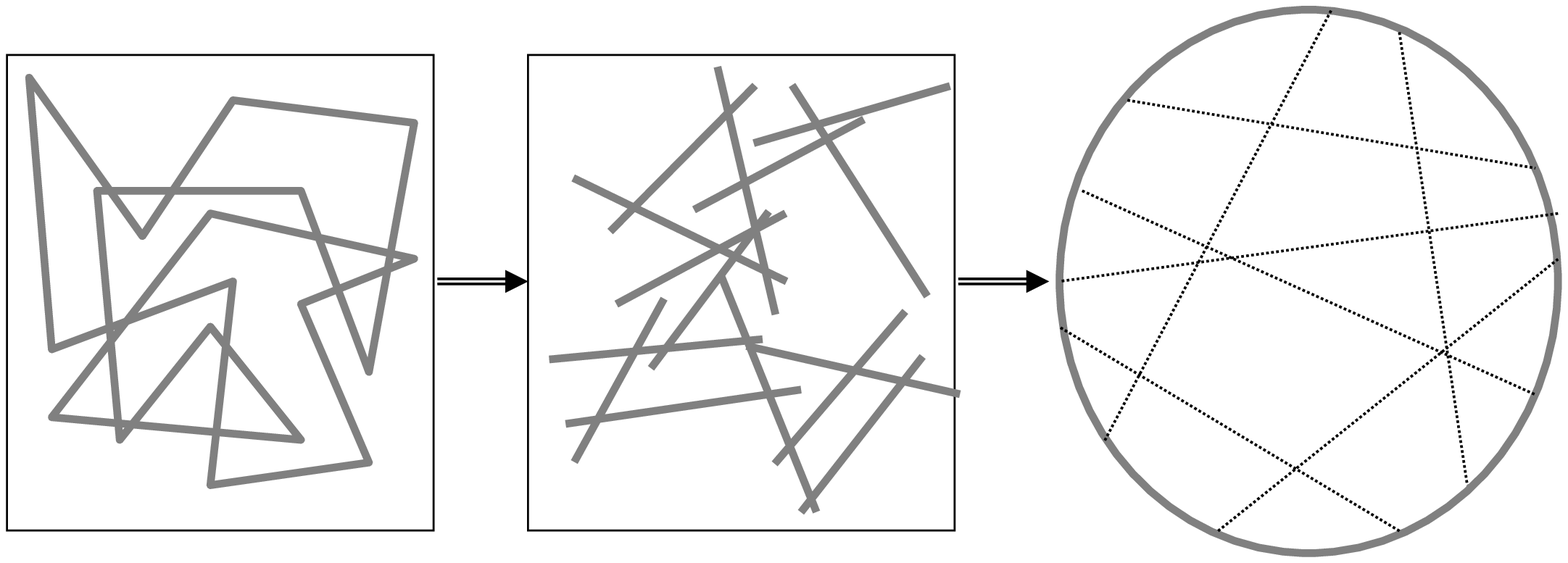}
\end{center}
\caption{Illustrated schematically is the simplified treatment of the folded
DNA. We first represent the DNA as the ideal gas of rods each with of a length
of one persistence length. Then we connect the rods randomly to form a small
world network (see text for details). Numerically we find the description to
work well.}%
\label{fig2}%
\end{figure}

To do so, we consider a packed DNA as an ideal gas of $\frac{L}{L_{0}}%
$\ straight rods of length $L_{0}$ that are distributed randomly in the cell
(see Fig. \ref{fig2}). The probability $p_{seg}$, that two given rods cross
within a distance of $R$ from each other is given by
\begin{equation}
p_{seg}=A\frac{L_{0}^{3}}{\Lambda^{3}}\frac{R^{2}}{L_{0}^{2}}=A\frac
{L_{0}R^{2}}{\Lambda^{3}}\;, \label{pseg}%
\end{equation}
where $A$ is a constant of order unity. Here $\frac{L_{0}^{3}}{\Lambda^{3}}$
is the probability that a given segments is located within a distance $L_{0}$
of a point inside the cell and $\frac{R^{2}}{L_{0}^{2}}$ is proportional to
the probability that this segment crosses a sphere of radius $R$ around the
point. Under the conditions described above we find that typically $p_{seg}%
\ll1$. Finally, to relate $p$ to $p_{seg}$ we note that to make an IT at least
one segment should be accessible. This yields
\begin{equation}
p=1-\left(  1-p_{seg}\right)  ^{L/L_{0}}\simeq1-e^{-A\frac{LR^{2}}{\Lambda
^{3}}}\;. \label{pgen}%
\end{equation}

Eq. (\ref{pgen}) implies that there are two possible regimes depending on the
value of $L$
\begin{equation}
p=\left\{
\begin{array}
[c]{c}%
A\frac{LR^{2}}{\Lambda^{3}}\ll1\hspace{0.5in}L\ll L_{c}\\
1\hspace{0.7in}\hspace{0.4in}L\gg L_{c}%
\end{array}
\right.  \;, \label{pGen}%
\end{equation}
where
\begin{equation}
L_{c}\sim\frac{\Lambda^{3}}{R^{2}}\;. \label{Lc}%
\end{equation}
In essence when $L\gg L_{c}$ (which can occur for example by having a large
protein) $p\simeq1$ and about half of the ITs are uncorrelated. However, when
$L\ll L_{c}$ we have that $p=A\frac{LR^{2}}{\Lambda^{3}}\ll1$, and most ITs
are correlated. The value of $L_{c}$ for the range of parameters of interest
is of the order of $10^{7}nm$\ for very large proteins ($R$ of order of tens
of $nm$, similar to the Lac repressor). Therefore \textit{in vivo} we expect a
relatively large $L_{c}$, so the regime $L\ll L_{c}$ should be
relevant\footnote{In a eucaryotic cell the concentration of DNA is much higher
and this statement may be wrong.}.

To summarize the above analysis we note that it effectively represents motion
on the DNA, using ITs, as motion on a one-dimensional discrete network. The
size of each site on this network is $b$, the scanned length on the DNA during
one binding event. Each step on the network takes on average a time $\delta$.
During an IT the protein can move from its position, $x$, to a randomly chosen
position in the interval $\left[  x-R,x+R\right]  $ along the chemical
coordinate (correlated transfer)\ with probability $1-p$ or to an uncorrelated
site with probability $p$ (uncorrelated transfer). Such networks are commonly
referred to as Small World Networks \cite{W99SM} (see Fig. \ref{fig2}(c)).

To find the relation between $\lambda$ and $p$ one has to consider the
dynamics of the DNA. Below we consider two extreme cases (a) a completely
quenched DNA configuration and (b) a strongly fluctuating DNA, which we term
annealed. A quenched DNA is static throughout the search process. An annealed
DNA changes its conformation on time scale much quicker than the motion of the protein.

\subsection{Quenched DNA\label{Quenched DNA}}

\begin{figure}[ptb]
\begin{center}
\includegraphics[
height=3.4575in, width=4.6216in ]{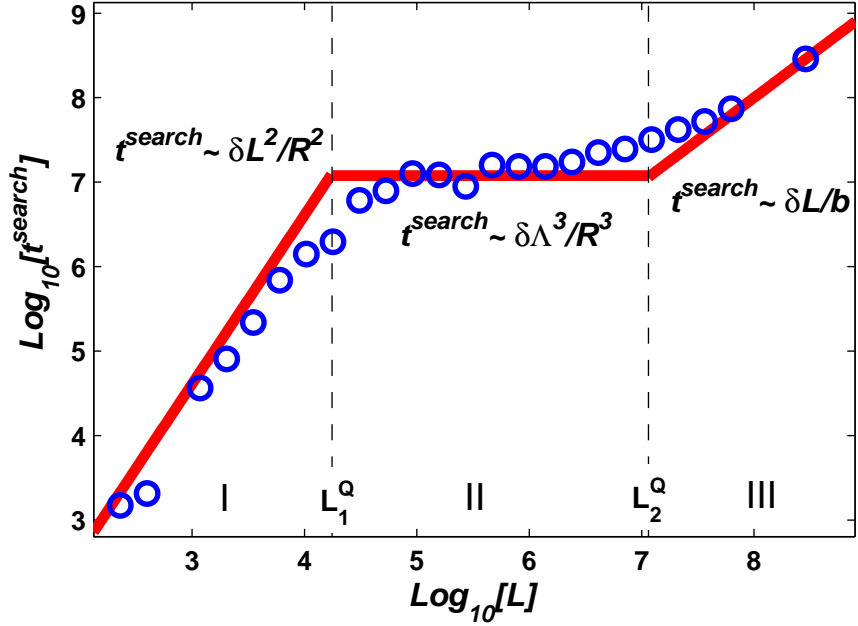}
\end{center}
\caption{The search time, $t^{search}$, is plotted as a function of $L$, the
DNA length for the pure IT case with a quenched DNA configuration. The circles
represent numerical data, while the solid line was obtained using Eqs.
(\ref{tQ1}) ,(\ref{tG}) and (\ref{tQ2}). The three visible regimes correspond
to the three on Fig. \ref{fig10} (see also Fig. \ref{Scheme}). In this plot
$R$ and $b$ were taken to be $3$ and $1$ lattice constants respectively (the
rest of the details are found in Appendix \ref{Appendix B}). The search time
is shown in units of $\delta$.}%
\label{fig4}%
\end{figure}

In this section we derive the search time for a quenched DNA. In particular we
will show that it is has a non-trivial behavior as a function of $L$. In the
regime that is expected to be relevant \textit{in vivo }the search time is
independent of the DNA's length (see Fig. \ref{fig4}).

\begin{figure}[ptb]
\begin{center}
\includegraphics[
height=2.0003in,
width=3in
]{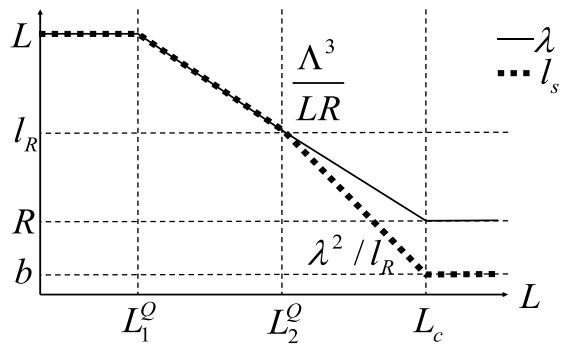}
\end{center}
\caption{The schematic behaviors of $\lambda$ and $l_{s}$ as a function of $L$
(on a log-log scale) is shown for quenched DNA and $b\ll R$.}%
\label{fig10}%
\end{figure}

For quenched DNA one expects that if an UIT can occur at point $x$ it can also
happen in a region of size $R$ around it. Similar considerations apply to
sites where an UIT can not occur. The typical distance traveled by the protein
along the DNA's chemical coordinate between two subsequent UITs, $L>\lambda
>R$, is of the order of the typical distance between two distinct locations
where an UIT can occur. This implies for $p\gg R/L$ a scaling of $\lambda$ of
the form
\begin{equation}
\lambda\sim\frac{R}{p}\;, \label{lamq}%
\end{equation}
where $p$ is defined above (see Eq. (\ref{pGen})) while for $p\ll R/L$ clearly
$\lambda=L$ (see Fig. \ref{fig10}).

From the previous discussion one may infer that there are three distinct
behaviors as a function of $L$ shown on Fig. \ref{fig10}. The first regime
occurs for DNA so short that an UIT cannot occur during the search. This
happens when $p\ll R/L$, or equivalently when \thinspace$L\ll L_{1}^{Q}$,
where using Eq. (\ref{pGen}) one finds
\begin{equation}
L_{1}^{Q}=\sqrt{\frac{\Lambda^{3}}{R}}\;.
\end{equation}
In fact, the estimate for $L_{1}^{Q}$ pushes the limit of our treatment since
the DNA is no longer densely packed in this regime. Nonetheless, we find good
agreement with numerical simulations.

The other regimes occur for $L\gg L_{1}^{Q}$, where one has $p\gg R/L$. In
this case the proteins can use UITs during the search. As discussed above
there is a length scale $L_{c}$ separating two distinct behaviors of $p$, and
therefore we have three different behaviors for $\lambda$ which are given by
(see Fig. \ref{fig10}):
\begin{equation}
\lambda\sim\left\{
\begin{array}
[c]{cc}%
L & \;\;\;\;L\ll\sqrt{\frac{\Lambda^{3}}{R}}=L_{1}^{Q}\\
\frac{\Lambda^{3}}{LR} & \;\;\;\;L_{1}^{Q}\ll L\ll L_{c}\\
R & \;\;\;\;\;\;L\gg\frac{\Lambda^{3}}{R^{2}}=L_{c}%
\end{array}
\right.  \;, \label{lamq2}%
\end{equation}
where as before $L_{c}=\Lambda^{3}/R^{2}$. Furthermore, as described above,
the scan between two subsequent UIT can either be recurrent ($\lambda\gg
l_{R}$) or transient ($\lambda\ll l_{R}$) with a crossover length $L_{2}^{Q}$.
This length scale $L_{2}^{Q}$ is determined by the condition $\lambda\left(
L=L_{2}^{Q}\right)  \sim l_{R}$. In the recurrent regime the walk between two
ITs doesn't skip locations on the DNA. This is in contrast to the transient
regime where many sites are skipped. Thus using Eqs. (\ref{pGen}) and
(\ref{lamq}) one finds
\begin{equation}
L_{2}^{Q}=\frac{\Lambda^{3}}{R^{3}}b\;. \label{L2Q}%
\end{equation}
For $L\gg L_{2}^{Q}$ the search between two subsequent UITs is short and
therefore transient while for $L\ll L_{2}^{Q}$ the search between two
subsequent UITs is long and therefore recurrent.

Note that when the search is transient, $t^{search}$ is independent of
$\lambda$ (see Eq. (\ref{1Gen})). Therefore, the crossover between two
distinct scaling behaviors of $t^{search}$ is governed by the smaller of the
two length scales $L_{c}$ and $L_{2}^{Q}$. For proteins performing only ITs
one expects $b$ to be smaller than $R$. It is easy to see that in such cases
$L_{2}^{Q}$ is smaller than $L_{c}$. (Other possibilities are discussed in
Sec. \ref{Intersegmental transfer and sliding}.)

To summarize there are two length scales $L_{1}^{Q}$ and $L_{2}^{Q}$ which
separate three possible regimes (see Fig. \ref{fig10}).

\begin{itemize}
\item Regime $\mathbf{I}$: $L\ll L_{1}^{Q}$
\end{itemize}

In this regime $\lambda\sim L$. There are no UITs and Eqs. (\ref{2Gen}) and
(\ref{1Gen}) give
\begin{equation}
t^{search}\sim\frac{L^{2}}{2D_{1eff}}\sim\frac{L^{2}}{R^{2}}\delta\;.
\label{tQ1}%
\end{equation}
This regime is clearly not relevant \textit{in vivo} (using the typical
values, $\Lambda\sim1\mu m$ and $R\sim10nm$, we find $L_{1}^{Q}\sim10\mu m$
which is much shorter than typical DNA lengths).

\begin{itemize}
\item Regime $\mathbf{II}$: $L_{1}^{Q}\ll L\ll L_{2}^{Q}$
\end{itemize}

Now the motion between two subsequent UITs is recurrent, $l_{R}\ll\lambda$,
and Eq. (\ref{lamq}) gives
\begin{equation}
\lambda\sim\frac{\Lambda^{3}}{LR}\;.
\end{equation}
Using Eqs. (\ref{2Gen}) and (\ref{1Gen}) we obtain%

\begin{equation}
t^{search}\sim\frac{\Lambda^{3}}{R^{3}}\delta\;. \label{tG}%
\end{equation}
Note that in this regime, as opposed to Sec. \ref{Sliding and jumping}, the
search time is \textit{independent} of the DNA's length. Eq. (\ref{tG}) is
equivalent to Eq. (\ref{t3D}) with an effective three-dimensional diffusion
coefficient $D_{3}\sim\frac{R^{3}}{r\delta}$. In contrast to the simple
three-dimensional diffusive search Eq. (\ref{tG}) does not depends on the
target size $r$ but rather on the protein size which may be much larger.

\begin{itemize}
\item Regime $\mathbf{III}$: $L\gg L_{2}^{Q}$
\end{itemize}

Here $\lambda\ll l_{R}$. and Eqs. (\ref{2Gen}) and (\ref{1Gen}) give%
\begin{equation}
t^{search}\sim\frac{L\delta}{b}\;. \label{tQ2}%
\end{equation}

The obtained results, compared to numerics, are summarized in Figs. \ref{fig4}
(see also Fig. \ref{Scheme}). One can clearly see the three regimes arising
for different lengths of DNA which are separated by $L_{1}^{Q}$ and $L_{2}%
^{Q}$. The details of the numerical simulation are described in Appendix
\ref{Appendix B}. Note that $L_{1}^{Q}$ and $L_{2}^{Q}$ are well predicted by
the scaling arguments.

The most relevant regime for \textit{in vivo} experiments in prokaryotic
organisms is likely to be the intermediate regime ($\mathbf{II}$) where the
search time is independent of the DNA's length and scales as $\Lambda^{3}$.
Comparing the search time in this regime (\ref{tG}) with the minimal search
time in the case when sliding and jumping are used, Eq. (\ref{tSM}), one may
see that if $\delta<R^{3}\sqrt{\frac{L}{\Lambda^{3}D_{1}D_{3}}}$ the search
time in the pure IT scenario is in fact smaller than the one of Sec.
\ref{Sliding and jumping} which includes only sliding and jumping case. This
is despite the fact that the protein \textit{never unbinds from the DNA}.

\subsection{Annealed DNA\label{Annealed DNA}}

In this section we consider the annealed case. As we show, here the search
time also has non-trivial but different than the quenched case behavior as a
function of $L$. In the regime that is expected to be relevant \textit{in vivo
}the search time scales as $\sqrt{L}$.

In the annealed case the time scale for a rearrangement of the DNA's
configuration is assumed to be much smaller than the time of the protein's
motion during an IT. As a result of the constant rearrangement of the DNA UITs
now occur with probability $p$ for each IT. The average number of ITs with no
UITs performed is therefore of the order of $\frac{1}{p}$ and thus the average
time that the protein spends between two subsequent UITs is $\frac{\delta}{p}$
($\delta$ as before is the typical time between two subsequent ITs). On
one-dimensional length scales smaller than $\lambda$ the protein diffuses with
a diffusion constant $D_{1eff}\sim\frac{R^{2}}{\delta}$. Therefore, the
typical one-dimensional distance between two subsequent UITs $\lambda$ is
\begin{equation}
\lambda\sim\sqrt{D_{1eff}\frac{\delta}{p}}\simeq\left\{
\begin{array}
[c]{cc}%
\sqrt{\frac{\Lambda^{3}}{L}} & \;\;\;\;L\ll L_{c}\\
R & \;\;\;\;L\gg L_{c}%
\end{array}
\right.  \;, \label{lama}%
\end{equation}
where $L_{c}$ is defined in Eq. (\ref{Lc}). As for the quenched case we will
see that again three distinct behaviors arise with two crossover lengths.

The first occurs when no UITs occur. The crossover length $L_{1}^{A}$ can be
extracted using the condition $\lambda\left(  L=L_{1}^{A}\right)  \sim L$
which under our assumptions on the protein's size can only occur when $L\ll
L_{c}$. This yields
\[
L_{1}^{A}\sim\Lambda\;.
\]
It is easy to see that $L_{1}^{A}\ll L_{1}^{Q}$. This means that, as expected,
in the annealed case the effects of UITs become important at much smaller DNA
concentration than in the quenched case. This happens because fast DNA
movements increase the probability to perform an UIT. As for $L_{1}^{Q}$, the
estimate for $L_{1}^{A}$ pushes the limit of our treatment since the DNA is no
longer densely packed in this regime.

\begin{figure}[ptb]
\begin{center}
\includegraphics[
height=2.0003in,
width=3in
]{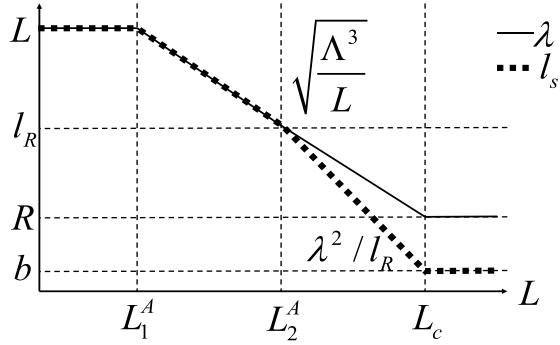}
\end{center}
\caption{The schematic behavior of $\lambda$ and $l_{s}$ as a function of $L$
(on a log-log scale) is shown for annealed DNA and $b\ll R$.}%
\label{fig11}%
\end{figure}

The second crossover length $L_{2}^{A}$ occurs when the motion between UITs
becomes transient. It can therefore be estimated using $\lambda\left(
L=L_{1}^{A}\right)  \sim l_{R}$. Taking the regime $L\ll L_{c}$ in Eq.
(\ref{lama}) yields
\begin{equation}
L_{2}^{A}\sim\frac{b^{2}\Lambda^{3}}{R^{4}}\;.
\end{equation}
For target sizes much smaller than the protein size ($b\ll R$), it is clear
that $L_{2}^{A}\ll L_{c}$ (see Eq. (\ref{Lc})). Hence, using the same
arguments as before, only two length scales, $L_{1}^{A}$ and $L_{2}^{A}$,
determines three possible regimes (see Fig. \ref{fig11}).

The three regimes which arise are:

\begin{itemize}
\item Regime $\mathbf{I}$: $L\ll L_{1}^{A}$
\end{itemize}

Here $\lambda\sim L$. There are no UITs and Eqs. (\ref{2Gen}) and (\ref{1Gen})
give
\begin{equation}
t^{search}\sim\frac{L^{2}}{2D_{1eff}}\sim\frac{L^{2}}{R^{2}}\delta\;.
\end{equation}

\begin{itemize}
\item Regime $\mathbf{II}$: $L_{1}^{A}\ll L\ll L_{2}^{A}$
\end{itemize}

Here searches between two subsequent UIT are recurrent so that $l_{R}%
\ll\lambda$. Eq. (\ref{lama}) gives%
\begin{equation}
\lambda\sim\sqrt{\frac{\Lambda^{3}}{L}}\;.
\end{equation}
Using Eqs. (\ref{2Gen}) and (\ref{1Gen}) we obtain%

\begin{equation}
t^{search}\sim\frac{\sqrt{L\Lambda^{3}}}{R^{2}}\delta\;. \label{tA}%
\end{equation}
Here, in contrast to the quenched case the intermediate result scales with the
length of the DNA as $L^{1/2}$. Note that the search time is always shorter
than that on a quenched DNA. This happens because the DNA's movement destroys
the correlation in the motion of the protein and, therefore, increases the
efficiency of the search. A similar dependence on $L$ ($t^{search}\propto
\sqrt{L}$) was obtained for a different model \cite{HS2007}. There, however,
the origin of the dependence is different, and is linked to modeling the DNA's
motion as diffusion of an ideal gas of rods.

\begin{itemize}
\item Regime $\mathbf{III}$: $L\gg L_{2}^{A}$
\end{itemize}

Here $\lambda\ll l_{R}$. Therefore, Eqs. (\ref{2Gen}) and (\ref{1Gen}) give%
\begin{equation}
t^{search}\sim\frac{L\delta}{b}\;.
\end{equation}
The obtained results are summarized later in Fig. \ref{Scheme}.

The most relevant regime for \textit{in vivo} experiments in prokaryotic
organisms is likely to be the intermediate one ($\mathbf{II}$) where the
search time scales as $L^{1/2}$ or alternatively as $\Lambda^{3/2}$. Comparing
the search time in this regime (\ref{tA}) with the minimal search time in the
case when sliding and jumping are used (\ref{tSM}) one may see that if
$\delta<\frac{R^{2}}{\sqrt{D_{1}D_{3}}}$ the search time in the pure IT case
is smaller than the one in the sliding and jumping case. This is despite the
fact that the protein \textit{never unbinds from the DNA}.

Numerical simulation of the annealed case require dynamical moves for the
whole DNA molecule. This is a formidable task for DNAs with reasonable length
which is beyond the scope of this paper.

\section{Intersegmental transfer and
sliding\label{Intersegmental transfer and sliding}}

Next we consider a protein that can perform both ITs and sliding. Namely, in
addition to ITs the protein can perform one-dimensional diffusion with only
one binding domain bound (see Fig. \ref{fig1}(a)). In the language of Sec.
\ref{Pure intersegmental transfer}, $b$ is now the typical sliding length
between two subsequent ITs. Now each step (distinct from a round defined
above), defined as the interval between the ends of two subsequent ITs, takes
a typical time $\delta=\frac{b^{2}}{2D_{1}}+\tau_{IT}$, where $D_{1}$ is the
one dimensional diffusion coefficient of the protein with only one binding
domain bound\footnote{The one-dimensional diffusion on the length scales
larger than $b$ has a different effective diffusion coefficient due to the
possibility of a CIT. Thus, to measure $D_{1}$ on large length-scales one
should not allow for ITs. This may by done, for example, by measuring the
motion of the part of the protein that contains only one binding domain
\cite{WAC2006,ELX2007}.} and $\tau_{IT}$ is the typical time that the protein
is bound to two DNA segments.

If $b\ll R$ it is straightforward to see that the results of the Sec.
\ref{Pure intersegmental transfer} hold with a redefined $\delta$. However, in
general the sliding length $b$ might be much larger than the proteins size
$R$. This is the regime that we focus on in this section.

Clearly, now the search between two subsequent UIT is always recurrent so that
$l_{s}\sim\lambda$. Here as before $\lambda$ is the typical distance traveled
by the protein between two subsequent UITs. However, now $D_{1eff}$ $\sim
\frac{b^{2}}{\delta}$, where as above $\delta=\frac{b^{2}}{2D_{1}}+\tau_{IT}$.
The search time as a function of $\lambda$, similar to Eq. (\ref{1Gen}),
becomes
\begin{equation}
t^{search}\sim\frac{L}{\lambda}\frac{\lambda^{2}}{D_{1eff}}\sim\frac{L\lambda
}{b^{2}}\delta\;. \label{bbR}%
\end{equation}

The value of $\lambda$, as in the previous section, depends on the dynamics of
the DNA molecule. Again we consider two extreme cases (a) quenched DNA and (b)
annealed DNA.

\subsection{Quenched DNA\label{Quenched DNA2}}

To obtain $\lambda$ we first introduce a new quantity, $\lambda_{0}$, defined
as the typical chemical distance between two locations in which the protein
\textit{can} perform an UIT. Note that we are interesting in the regime $b\gg
R$. Therefore the values of $\lambda$ and $\lambda_{0}$ may be distinct since
an UIT is not necessarily performed at every possible location on the DNA.
Clearly, however, the functional behavior of $\lambda_{0}$ is identical to
that of $\lambda$ in the previous section. This yields (see Eq. \ref{lamq2})
\begin{equation}
\lambda_{0}\sim\left\{
\begin{array}
[c]{cc}%
L & \;\;\;\;L\ll\sqrt{\frac{\Lambda^{3}}{R}}=L_{1}^{Q}\\
\frac{\Lambda^{3}}{LR} & \;\;\;\;L_{1}^{Q}\ll L\ll L_{c}\\
R & \;\;\;\;\;\;\;\;L\gg\frac{\Lambda^{3}}{R^{2}}=L_{c}\;
\end{array}
\right.  \;, \label{lam00}%
\end{equation}
where we have used the definitions of $L_{1}^{Q}$ and $L_{c}$ of the previous section.

\begin{figure}[ptb]
\begin{center}
\includegraphics[
height=2.0003in,
width=3in
]{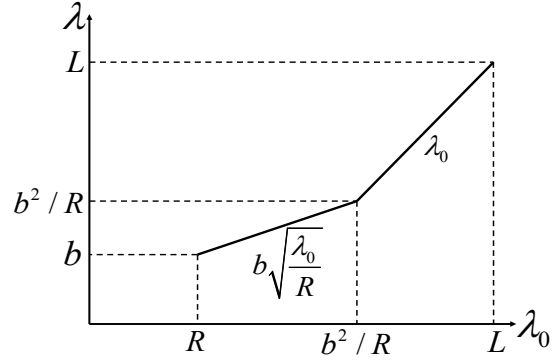}
\end{center}
\caption{The schematic behavior of $\lambda$ and $l_{s}$ as a function of $L$
(on a log-log scale) is shown for quenched DNA, $L\gg\frac{b^{2}}{R}$ and
$b\gg R$.}%
\label{fig8}%
\end{figure}

Similar to the derivation of Eq. (\ref{lR}), when $\lambda_{0}/b\gg b/R$, the
effective random walk of the protein along a length $\lambda_{0}$ is
recurrent. Here recurrent motion implies that sites where an UIT can occur are
visited many times before a neighboring site where an UIT can occur is met
(note that this is distinct from the recurrent behavior of Sec.
\ref{Pure intersegmental transfer}). In the recurrent regime a location of a
possible UITs is visited many times and therefore not missed. In this case
$\lambda\sim\lambda_{0}$. In the opposite transient regime (again distinct in
meaning from that used in Sec. \ref{Pure intersegmental transfer}),
$\frac{\lambda_{0}}{b}\ll\frac{b}{R}$ and the protein performs an UIT only
after it travels a distance $\lambda\gg\lambda_{0}$. In the latter regime each
IT has a probability $\frac{R}{\lambda_{0}}$ to be an UIT. Therefore between
two subsequent UITs the protein performs $\frac{\lambda_{0}}{R}$ ITs. Using
the diffusive nature of the motion we find $\lambda\sim b\sqrt{\frac
{\lambda_{0}}{R}}$. The value of $\lambda$ as function of $\lambda_{0}$ is
shown schematically in Fig. \ref{fig8}.

Combining the three regimes of $\lambda_{0}$ with the above mentioned
crossover from $\lambda\sim\lambda_{0}$ to $\lambda\sim b\sqrt{\frac
{\lambda_{0}}{R}}$ (which occurs at $L=\Lambda^{3}/b^{2}$) one finds, using
$b/R\gg1$, \textit{four} regimes for the search time:

\begin{itemize}
\item Regime $\mathbf{I}$ occurs for $L\ll L_{1}^{Q}$ corresponding to
$\lambda\sim\lambda_{0}=L$ in Eq. (\ref{lam00}). Using Eq. (\ref{bbR}) gives
\begin{equation}
t^{search}\sim\frac{L^{2}}{b^{2}}\delta\;. \label{1Q}%
\end{equation}

\item Regime $\mathbf{II}$ occurs for $\frac{\Lambda^{3}}{b^{2}}\gg L\gg
L_{1}^{Q}$ and $\lambda\sim\lambda_{0}\sim\frac{\Lambda^{3}}{LR}$. Using Eq.
(\ref{bbR}) yields
\begin{equation}
t^{search}\sim\frac{\Lambda^{3}}{b^{2}R}\delta\;. \label{2Q}%
\end{equation}

\item Regime $\mathbf{III}$ occurs for $\frac{\Lambda^{3}}{b^{2}}\ll L\ll
L_{c}$. Now $\lambda\sim b\sqrt{\frac{\lambda_{0}}{R}}\sim b\sqrt
{\frac{\Lambda^{3}}{LR^{2}}}$. Using Eq. (\ref{bbR}) we find
\begin{equation}
t^{search}\sim\frac{\sqrt{\Lambda^{3}L}}{bR}\delta\;. \label{3Q}%
\end{equation}

\item Regime $\mathbf{IV}$ occurs for $L\gg L_{c}$. Here $\lambda\sim
b\sqrt{\frac{\lambda_{0}}{R}}\sim b$ and with Eq. (\ref{bbR}) one gets
\begin{equation}
t^{search}\sim\frac{L}{b}\delta\;. \label{4Q}%
\end{equation}

\end{itemize}

\begin{figure}[ptb]
\begin{center}
\includegraphics[
height=4in, width=5.33in ]{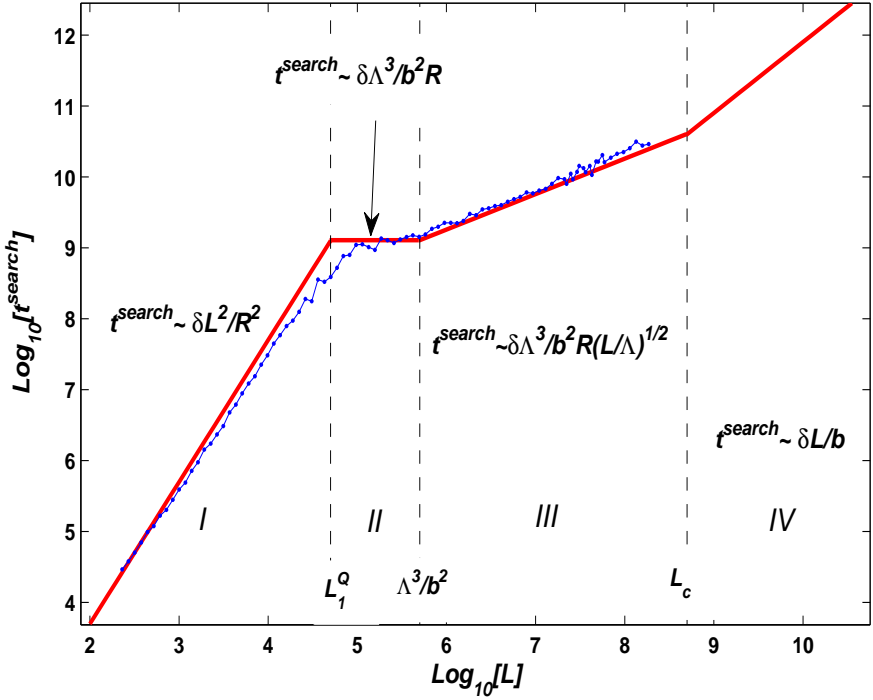}
\end{center}
\caption{ \ $t^{search}$ is plotted as a function of $L$, the DNA length for a
model with IT and sliding for a quenched DNA. The thin line with dots
represent numerical data, while the bold solid line was obtained using Eqs.
(\ref{1Q}) ,(\ref{2Q}), (\ref{3Q}) and (\ref{4Q}) (see also Fig.
\ref{Scheme}). In this plot $R$ and $b$ were taken to be $1$ and $20$ lattice
constants respectively (the rest of the details could be found in Appendix
\ref{Appendix B}). The search time is shown in units of $\delta$. }%
\label{ITwithS}%
\end{figure}

Fig. \ref{ITwithS} shows a comparison between the four theoretically predicted
regimes and the numerical simulation of the model. Three regimes are
reproduced by the numerics while the fourth one was not reproduced due to
computational limitations.

For a moderate values of $\tau_{IT}$ one may see that long sliding may
drastically decrease the efficiency of the search. This occurs because long
sliding prevents both UITs that destroy correlations in the search process and
CITs that increase the effective one-dimensional diffusive constant.

\subsection{Annealed DNA\label{Annealed DNA2}}

Here using the arguments presented in Sec. \ref{Annealed DNA}, the average
number of steps performed between two subsequent UITs is of the order of
$\frac{1}{p}$ where $p$ is given in Eq. (\ref{pGen}). This implies a typical
time between the subsequent UITs of the order of $\frac{\delta}{p}$. Using the
fact that along the DNA the motion of the protein is diffusive with an
effective diffusion constant $D_{1eff}\sim\frac{b^{2}}{\delta}$ one finds
\begin{equation}
\lambda\sim\sqrt{D_{1eff}\frac{\delta}{p}}. \label{lamA}%
\end{equation}
Clearly, $\lambda$ can only take values in the range $b\leq\lambda\leq L$.
These with the possible values of $p$ (see Eq. (\ref{pGen})) define the
borders of the following three regimes:

\begin{itemize}
\item Regime $\mathbf{I}$ occurs for $\lambda\sim L$. Using Eq. (\ref{lamA})
and $p=LR^{2}/\Lambda^{3}$ it can be verified that this regime occurs when
$L\ll\Lambda\left(  \frac{b}{R}\right)  ^{2/3}$. In this case no UIT occur
during the search and Eq. (\ref{bbR}) gives
\begin{equation}
t^{search}\sim\frac{L^{2}}{b^{2}}\delta\;.
\end{equation}

\item Regime $\mathbf{II}$ occurs when $\Lambda\left(  \frac{b}{R}\right)
^{2/3}\ll L\ll L_{c}$. Using Eq. (\ref{lamA}) and $p=LR^{2}/\Lambda^{3}$ one
finds that in this case $\lambda\sim b\sqrt{\frac{\Lambda^{3}}{LR^{2}}}$.
Using Eq. (\ref{bbR}) gives
\begin{equation}
t^{search}\sim\frac{\sqrt{\Lambda^{3}L}}{bR}\delta\;.
\end{equation}

\item Regime $\mathbf{III}$ occurs where $L\gg L_{c}$ and almost all ITs are
UITs. Here $\lambda\sim b$ and $p\simeq1$ so that Eq. (\ref{bbR}) gives
\begin{equation}
t^{search}\sim\frac{L}{b}\delta\;.
\end{equation}

\end{itemize}

The obtained results are summarized in Fig. \ref{Scheme}.

One may see that in the case of long sliding, rapid DNA motion cannot decrease
the search time significantly as in the pure IT case. This is because long
sliding prevents fast decay of correlations.

\subsection{Motion with no CIT\label{Motion with no CIT}}

Here we consider a case where the structure of the protein causes it to prefer
UITs over CITs. This may occur, for example, in cases where the
\textquotedblleft legs\textquotedblright\ of the protein are antiparallel and
rigid. The motion on length scale smaller than $\lambda$ is then diffusive
involving only sliding with a diffusion coefficient $D_{1}$. In this case,
clearly $l_{s}=\lambda$ and the time between two subsequent UITs is given
$\frac{\lambda^{2}}{2D_{1}}+\tau_{IT}$ where $\tau_{IT}$ is the time of an
UIT. One finds, similar to Sec. \ref{Sliding and jumping}
\begin{equation}
t^{search}\sim\frac{L}{\lambda}\left(  \frac{\lambda^{2}}{2D_{1}}+\tau
_{IT}\right)  \;. \label{tnoCIT}%
\end{equation}
The relationship between this and the picture of Sec.
\ref{Sliding and jumping} is given by identifying the antenna's length $l$
with $\lambda$ and the three-dimensional diffusion time $\tau_{3}$ with
$\tau_{IT}$.

\bigskip\bigskip

\begin{figure}[ptb]
\begin{center}
\includegraphics[
height=3.5in, width=7in ] {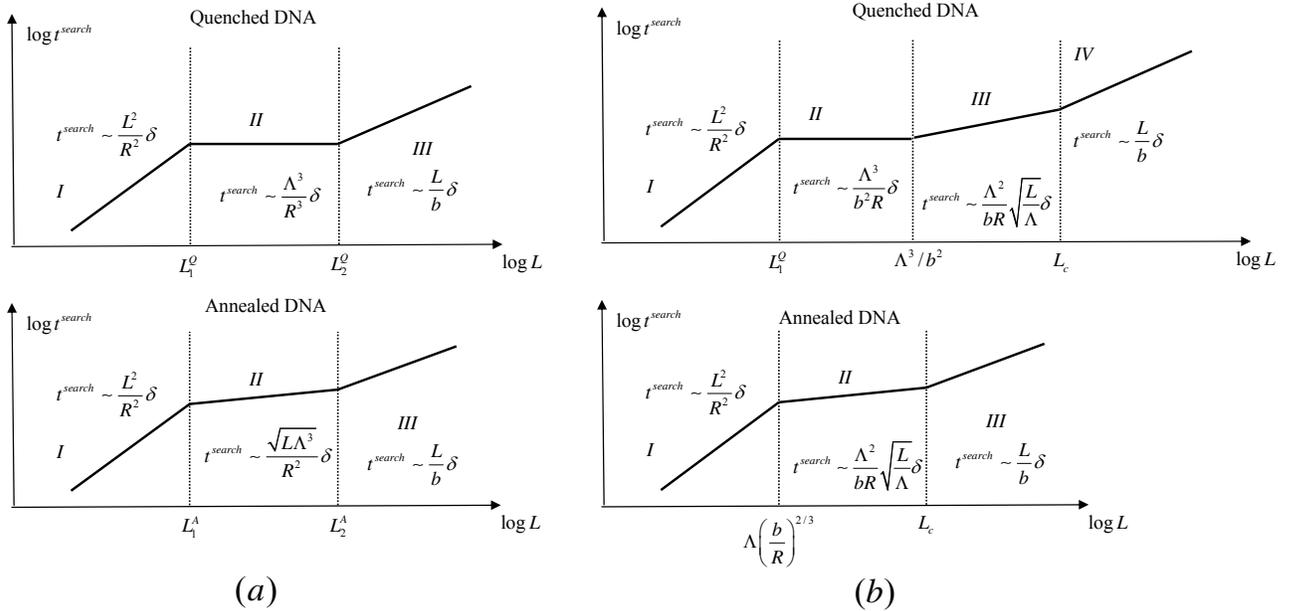}
\end{center}
\caption{ \ In this figure the schematic behavior of $t^{search}$ as a
function of the DNA length $L$ is shown in absence of the jumps. (a) shows
short sliding results ($b\ll R$). (b) shows long sliding results ($b\gg R$).}%
\label{Scheme}%
\end{figure}

Most of the results of Secs. \ref{Pure intersegmental transfer} and
\ref{Intersegmental transfer and sliding} are summarized in Fig. \ref{Scheme}.
The results of this section indicate that ITs may supply reasonable search
times if they are quick enough. Combining IT with sliding we see that even
rare UIT events may break correlations created by one-dimensional diffusion.
In this sense ITs act as jumps without the need for detachment from the DNA.
Besides this, CITs may effectively accelerate the one-dimensional diffusion or
even replace it altogether.

\section{Intersegmental transfer and
jumping\label{Intersegmental transfer and jumping}}

We now turn to consider the effect of jumping on the results described above.
Before addressing the full problem, including ITs sliding and jumping, we
first consider a model in which only ITs and jumps occur, and ignore sliding.
To include jumping we assign a probability $\frac{dt}{\tau_{1}}$ for a protein
to detach from the DNA during a time interval $dt$. The unbinding initiates a
jump in which the protein uses three-dimensional diffusion to rebind at a new
location on the DNA. Note that since there is no sliding it is safe to assume
$b\ll R$.

As argued in the previous section, it is reasonable that both UITs and jumps
move the protein to a new location which is chosen randomly on the DNA.
Therefore, the search process is composed of a series of one-dimensional scans
(occurring through CITs) of the DNA interrupted by uncorrelated relocations.
The uncorrelated relocations can occur through two independent processes:
jumps and UITs. The typical search time can be evaluated using an approach
identical to that of the previous sections.

First, we need to estimate the typical time $\tau_{1eff}$ between two
uncorrelated relocations. Combining, the previously derived typical time
between two subsequent UITs, $\frac{\lambda^{2}}{2D_{1eff}}$, and the typical
time between jumps $\tau_{1}$ we obtain\footnote{This expression is exact in
the annealed case but it is only an approximation in the quenched regime.
However, the error does not exceed $50\%$ (see Appendix \ref{D1} for
details).}
\begin{equation}
\tau_{1eff}\simeq\frac{1}{\frac{2D_{1eff}}{l^{2}}+\frac{2D_{1eff}}{\lambda
^{2}}}\;, \label{t1eff}%
\end{equation}
where $\lambda$, defined before, is the typical distance that the protein
travels between two subsequent UITs and we define an antenna length
$l=\sqrt{2D_{1eff}\tau_{1}}$.

Here and in the next section we focus on the search time as a function of $l$.
This quantity is influenced by the protein-DNA non-specific binding energy and
governs the frequency of jumps. Other parameters that do not depend on $l$,
such as $\lambda$, are taken as fixed. The value of $\lambda$ relevant for the
discussion here is given in Sec. \ref{Pure intersegmental transfer} , where
$b\ll R$. Note, that when incorporated in the results below the resulting
behavior is very complicated. While this is easy to obtain we skip all the
regimes and focus on important qualitative behavior.

To proceed we note that the typical distance between two uncorrelated
relocation events is given by
\begin{equation}
l_{eff}=\sqrt{2D_{1eff}\tau_{1eff}}\simeq l\sqrt{\frac{1}{1+\frac{l^{2}%
}{\lambda^{2}}}}\;.\label{leff}%
\end{equation}
As expected, and seen in Eqs. (\ref{t1eff}) and (\ref{leff}), the relative
importance of both mechanisms is controlled by the ratio $\frac{l}{\lambda}$.
In the case of $l/\lambda\gg1$ jumping is rare compared to UITs and may be
neglected leading to the behavior found in Sec.
\ref{Pure intersegmental transfer}. In the opposite case $l/\lambda\ll1$ the
possibility of performing an UIT is negligible and the results of Sec.
\ref{Sliding and jumping} hold.

Finally, we must estimate the average time spent by the protein performing one
uncorrelated relocation. This is given by the average of the jump time,
$\tau_{3}$, and the time of an IT, weighed with the probability of performing
each. This gives
\begin{align}
\tau_{3eff} &  =\tau_{3}\frac{\tau_{1eff}}{\tau_{1}}+\delta\left(
1-\frac{\tau_{1eff}}{\tau_{1}}\right)  \nonumber\\
&  =\tau_{3}\frac{l_{eff}^{2}}{l^{2}}+\delta\left(  1-\frac{l_{eff}^{2}}%
{l^{2}}\right)  \simeq\frac{\tau_{3}+\delta\frac{l^{2}}{\lambda^{2}}}%
{1+\frac{l^{2}}{\lambda^{2}}}\;,\label{t3eff}%
\end{align}
where $\frac{1/\tau_{1}}{1/\tau_{eff}}$ is the probability of a jump,
$1-\frac{1/\tau_{1}}{1/\tau_{eff}}=\frac{1/\frac{\lambda^{2}}{2D_{1eff}}%
}{1/\tau_{eff}}$ is the probability of an UIT and $\delta$, defined above is
the time of an IT (see Appendix \ref{D2} for a more detailed derivation).

The total search time, as before, takes the form of Eq. (\ref{SM}). Now, each
search round is defined as the interval between two subsequent uncorrelated
relocations. The total time of one round is $\tau_{r}\sim\tau_{1eff}%
+\tau_{3eff}$, and therefore the search time is given by
\begin{equation}
t^{search}\sim N_{r}\tau_{r}\sim\frac{L}{l_{s}}\left(  \tau_{1eff}+\tau
_{3eff}\right)  \;. \label{tJIT}%
\end{equation}
Here $l_{s}$ is the length \textit{scanned} between two subsequent
uncorrelated relocations. In the case discussed here $b\ll R$, and the value
of $l_{s}$ depends on the properties of the search between two uncorrelated
relocations, namely the ratio of $l_{eff}$ and $l_{R}$, the recurrence length
(see Eq. (\ref{lR}) and the relevant discussion). If $l_{eff}\gg l_{R}$ the
search between two subsequent jumps is recurrent and $l_{s}\sim l_{eff}$.
However, in the opposite regime, $l\ll l_{R}$, $l_{s}\sim\frac{l_{eff}^{2}%
}{l_{R}}$.

\begin{figure}[ptb]
\begin{center}
\includegraphics[
height=2.6878in, width=4.0309in ]
{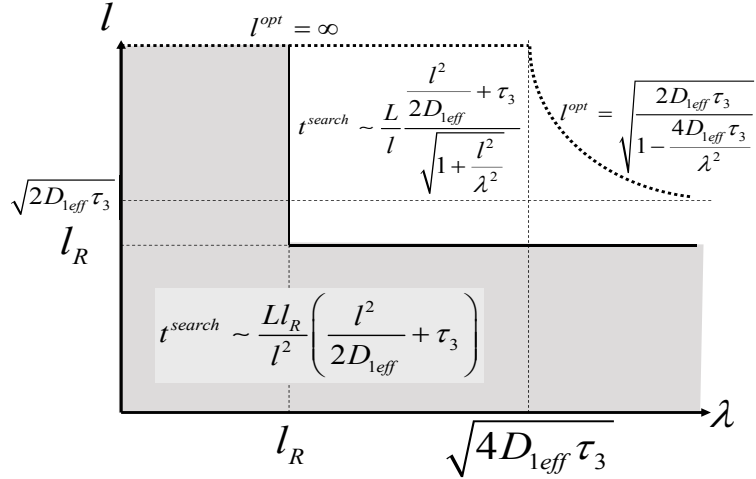}
\end{center}
\caption{Possible regimes as a function of $l$ and $\lambda$ are shown in the
case of ITs and jumping (or IT, jumping and sliding with $b\ll R$) for
$l_{R}\ll\sqrt{2D_{1eff}\tau_{3}}$. The gray (white) area represents regime
$\mathbf{I}$ ($\mathbf{II}$). The dashed line represents the optimal antenna
length. The optimal antenna length in the absence of IT is equal to
$\sqrt{2D_{1}\tau_{3}}$.}%
\label{fig6}%
\end{figure}

\begin{figure}[ptb]
\begin{center}
\includegraphics[
height=2.6809in, width=4.0205in ] {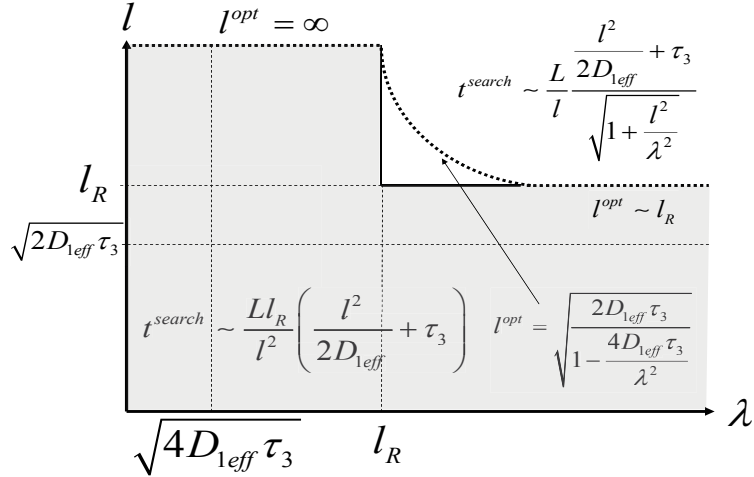}
\end{center}
\caption{Possible regimes as a function of $l$ and $\lambda$ are shown in the
case of ITs and jumping (or IT, jumping and sliding with $b\ll R$) for
$l_{R}\gg\sqrt{2D_{1eff}\tau_{3}}$. The gray (white) area represents regime
$\mathbf{I}$ ($\mathbf{II}$). The dashed line represents the optimal antenna
length. The optimal antenna length in the absence of IT is equal to
$\sqrt{2D_{1}\tau_{3}}$.}%
\label{fig7}%
\end{figure}

Therefore, for a given $\lambda$ there are two regimes (see Fig. \ref{fig6}
and \ref{fig7}):

\begin{itemize}
\item Regime $\mathbf{I}$ ($l_{R}\ll l_{eff}$):
\end{itemize}

In this regime, using Eq. (\ref{tJIT}), the total search time is
\begin{align}
t^{search}  &  \sim N_{r}\tau_{r}\sim\frac{L}{l_{eff}}\left(  \tau_{1eff}%
+\tau_{3eff}\right)  \sim\nonumber\\
&  \sim\frac{L}{l}\left(  \frac{\frac{l^{2}}{2D_{1eff}}+\tau_{3}+\delta
\frac{l^{2}}{\lambda^{2}}}{\sqrt{1+\frac{l^{2}}{\lambda^{2}}}}\right)
\simeq\frac{L}{l}\frac{\frac{l^{2}}{2D_{1eff}}+\tau_{3}}{\sqrt{1+\frac{l^{2}%
}{\lambda^{2}}}}\;, \label{tIJTres}%
\end{align}
where we used $\lambda\gg R$.

Comparing with Eq. (\ref{SM}) we note that here we have both an effective
diffusion constant and an extra \textit{\ enhancement factor} given by
$\left(  1+\frac{l^{2}}{\lambda^{2}}\right)  ^{-1/2}$. As we now show, this
factor has important consequence.

Consider the value of $\tau_{1}=\frac{l^{2}}{2D_{1eff}}$ for which a minimal
search time is obtained and compare it with the usual paradigm of $\left(
\tau_{1}^{opt}\right)  _{0}=\tau_{3}$. Due to the enhancement factor we now
find
\begin{equation}
\tau_{1}^{opt}=\frac{\left(  \tau_{1}^{opt}\right)  _{0}}{1-\frac
{4D_{1eff}\tau_{3}}{\lambda^{2}}}\;,\label{lopt}%
\end{equation}
where $\left(  \tau_{1}^{opt}\right)  _{0}=\tau_{3}$ (see Eq. (\ref{t1opt0}%
))\ is the optimal antenna size in absence of ITs ($\lambda\rightarrow\infty$)
(see Sec. \ref{Sliding and jumping}). It is interesting to note that $l^{opt}$
approaches infinity when $\tau_{3}$ is larger than a critical value
\begin{equation}
\tau_{3c}=\frac{\lambda^{2}}{4D_{1eff}}\;.\label{t3c}%
\end{equation}
Hence, the minimal search time for $\tau_{3}\geq\tau_{3c}$, is
\textit{identical to that with no jumps} (see Sec.
\ref{Pure intersegmental transfer}). It is important to note that $\tau_{3c}$
depends, as expected, on the time of an IT through $D_{1eff}$. In the case
when $\tau_{3}\leq\tau_{3c}$ Eqs. (\ref{tIJTres}) and (\ref{lopt}) give
\begin{equation}
t_{opt}^{search}\sim L\sqrt{\frac{\tau_{3}}{D_{1eff}}}\sqrt{1-\frac{\tau_{3}%
}{2\tau_{3c}}}\;.
\end{equation}
In this regime $t_{opt}^{search}$ is monotonically increasing in $\tau_{3}$.

\begin{figure}[ptb]
\begin{center}
\includegraphics[
height=3.0926in,
width=4.7184in
]{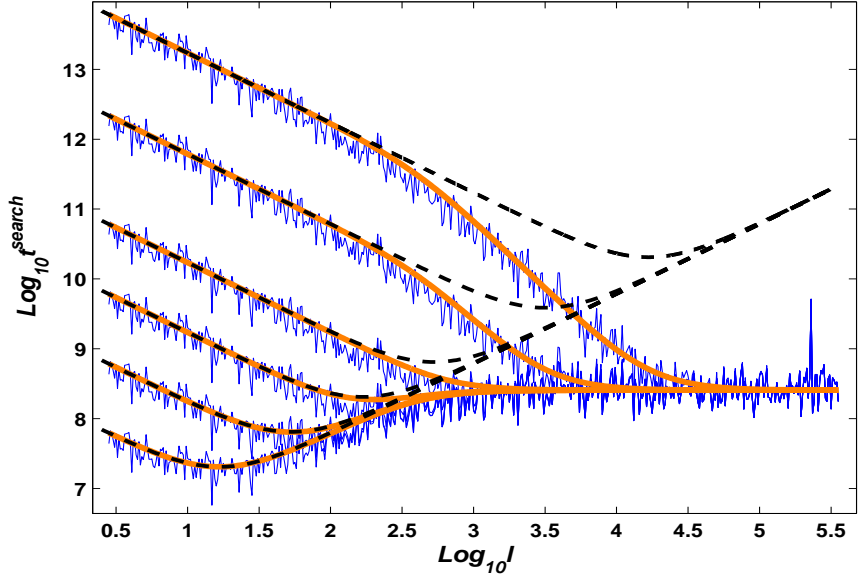}
\end{center}
\caption{The influence of ITs on the search time is shown.The search time,
$t^{search}$, is plotted as a function of the antenna length, $l,$ for a
different values of $\tau_{3}$
($140,\ 1400,\ 14000,\ 1400000,\ 5000000,\ 14000000$ in units of $\delta$ from
bottom up). Here only ITs and jumping are allowed. Thin solid lines represent
the numerical results. The bold solid lines represent analytic results (Eq.
(\ref{tIJTres})). The black, dashed lines represent the search time in the
case with no ITs, obtained by using Eq. (\ref{ts1}) with the effective
diffusion constant $D_{1eff}=\frac{R^{2}}{2\delta}$ instead of $D_{1}$. Here
$L$, $R$ and $b$ were taken to be $1224000$, $1$ and $1$ lattice constants
respectively. Since $R=b=1$ diffusion through sliding is identical to one
through CITs. This allows us to directly compare sliding and jumping with ITs
and jumping.}%
\label{fig5}%
\end{figure}

In Fig. \ref{fig5} we show a comparison between the results of numerical
simulation and Eq. (\ref{tIJTres}).

\begin{itemize}
\item Regime $\mathbf{II}$\ ($l_{eff}\ll l_{R}$)
\end{itemize}

In this case $l_{s}\sim\frac{l_{eff}^{2}}{l_{R}}$ and Eq. (\ref{tJIT}) yields
\begin{equation}
t^{search}\sim\frac{Ll_{R}}{l_{eff}^{2}}\left(  \tau_{1eff}+\tau
_{3eff}\right)  \sim\frac{Ll_{R}}{l^{2}}\left(  \frac{l^{2}}{2D_{1eff}}%
+\tau_{3}\right)  \;. \label{tIJTres2}%
\end{equation}
Interestingly, in this regime the minimal search time is obtained when
$\tau_{1}$ diverges ; This means that jumping \textit{only increase the search
slower in this case}. We note that some care needs to be taken with the limit
since if $\lambda>l_{R}$ and the value of $l$ exceeds $l_{R}$ the regime
$l_{eff}\ll l_{R}$ transforms into Regime $\mathbf{I}$.



The results of this section highlight several interesting features which will
also appear in the more general case, where sliding is also allowed. First, we
note that in the limit of very strong protein-DNA affinity (large values of
$\tau_{1}$) the search time becomes robust to changes in the value of
$\tau_{1}$. This is very different from a search process with no ITs (see Eqs.
(\ref{tIJTres}), (\ref{tIJTres2}) and Fig. \ref{fig5}), and may give a
possible explanation to the difference between \textit{in vitro} experiments
on the Lac repressor \cite{RBC70}. There a strong dependence of the search
time on ionic strength (and therefore on the protein-DNA affinity) was found.
However, \textit{in vivo} experiment \cite{RCMKAFR87} found that the
efficiency of the repression by the same protein is very robust to changes in
the ionic strength.

Furthermore, by examining the optimal search time, we find that beyond some
critical value of $\tau_{3}$ jumps increase the search time (see Fig.
\ref{fig5} for demonstration). This may give a possible explanation of the
obtained value of $\tau_{1}$ \textit{in vitro} \cite{WAC2006} and \textit{in
vivo} \cite{ELX2007} for the Lac repressor. These are much larger than the
optimal $\tau_{1}$ predicted by models that do not include ITs.

In Fig. \ref{fig5} a comparison between Eq. (\ref{tIJTres}) and numerical
simulation is shown. One may see that increasing the value of $\tau_{3}$
increases the optimal value of $l$ (or equivalently $\tau_{1}$) in such a way
that above some critical value, predicted by Eq. (\ref{t3c}), it becomes infinite.

\section{Intersegmental transfer, sliding and
jumping\label{Intersegmental transfer, sliding and jumping}}

With the results of the previous section it is straightforward to consider the
general case where ITs, sliding and jumping are allowed. Similar to the
previous section we show that jumping may slow the search process
significantly. However, ITs make the search process much more robust to
variations in parameters.

First consider the case $b\ll R$ where sliding events are very short. Clearly,
in this case the results of the previous section hold with $\delta=\frac
{b^{2}}{2D_{1}}+\tau_{IT}$. Here as in Sec.
\ref{Intersegmental transfer and sliding}, $D_{1}$ is the one dimensional
diffusion coefficient for sliding and $\tau_{IT}$ is the typical time that the
protein is bound to two DNA segments. With this in mind, in this section we
discuss only the opposite case of $b\gg R$. Here, as in Sec.
\ref{Intersegmental transfer and jumping}, the parameters that do not depend
on $l$, such as $\lambda$, are taken as given. In Sec.
\ref{Intersegmental transfer and sliding} contains the relevant derivation of
$\lambda$ is calculated for the case discussed here of long sliding, $b\gg R$.

As shown in Sec. \ref{Intersegmental transfer and sliding} in this case
$D_{1eff}$ $\sim\frac{b^{2}}{2\delta}$ with $\delta=\frac{b^{2}}{2D_{1}}%
+\tau_{IT}$. Following Sec. \ref{Intersegmental transfer and jumping} we first
need $\tau_{3eff}$, the typical time of an uncorrelated relocation. This is
given by (see the derivation of Eq. (\ref{t3eff}) and Appendix \ref{D1})
\begin{equation}
\tau_{3eff}=\frac{\tau_{3}+\tau_{IT}\frac{l^{2}}{\lambda^{2}}}{1+\frac{l^{2}%
}{\lambda^{2}}}\;.
\end{equation}

Note that here, since $b\gg R$, the search between two subsequent uncorrelated
relocations is always recurrent and therefore $l_{s}\sim l_{eff}$. Therefore,
similar to Sec. \ref{Intersegmental transfer and jumping}, the search time is
given by
\begin{equation}
t^{search}\sim\frac{L}{l_{eff}}\left(  \tau_{1eff}+\tau_{3eff}\right)
\sim\frac{L}{l_{eff}}\left(  \frac{l_{eff}^{2}}{2D_{1eff}}+\frac{\tau_{3}%
+\tau_{IT}\frac{l^{2}}{\lambda^{2}}}{1+\frac{l^{2}}{\lambda^{2}}}\right)  \;.
\label{tblR}%
\end{equation}
Using Eqs. (\ref{leff}) and (\ref{tblR}), the total search time can be written
as
\begin{equation}
t^{search}\sim\frac{L}{l\sqrt{1+\frac{l^{2}}{\lambda^{2}}}}\left(  \frac
{l^{2}}{2D_{1eff}}+\tau_{IT}\frac{l^{2}}{\lambda^{2}}+\tau_{3}\right)  \;.
\label{tblRs}%
\end{equation}

Again, it is interesting to consider the optimal value of $\tau_{1}$
\begin{equation}
\tau_{1}^{opt}=\frac{\left(  \tau_{1}^{opt}\right)  _{0}}{1-\frac{2\tau
_{3}-\tau_{IT}}{\lambda^{2}/2D_{1eff}}}\;,\label{loptSITJ}%
\end{equation}
where $\left(  \tau_{1}^{opt}\right)  _{0}=\tau_{3}$ (see Eq. (\ref{t1opt0}%
))\ is the optimal antenna size in absence of ITs ($\lambda\rightarrow\infty
$).

Interestingly, Eq. (\ref{loptSITJ}) shows that the optimal $\tau_{1}^{opt}$,
may either be smaller or larger than $\left(  \tau_{1}^{opt}\right)  _{0}$
depending on the time of an IT, $\tau_{IT}$. It is also noteworthy that when
$2\tau_{3}>\lambda^{2}/2D_{1eff}+\tau_{IT}$ the optimal $\tau_{1}$ value
becomes infinite. Namely, \textit{jumping makes the search process slower}.
This is similar to the behavior found in Sec.
\ref{Intersegmental transfer and jumping}, and again the critical value of
$\tau_{3}$ depends on microscopic quantities such as the time of an IT.

The minimal search time obtain is
\begin{equation}
t_{opt}^{search}\sim\left\{
\begin{array}
[c]{cc}%
\frac{L}{\lambda}\sqrt{\tau_{3}}\sqrt{\lambda^{2}/2D_{1eff}+\tau_{IT}-\tau
_{3}} & \;\;\;\;\;\;\;\tau_{3}<\frac{\lambda^{2}/2D_{1eff}+\tau_{IT}}{2}\\
L\left(  \frac{1}{2D_{1eff}}+\frac{\tau_{IT}}{\lambda^{2}}\right)   &
\;\;\;\;\;\;\;\tau_{3}>\frac{\lambda^{2}/2D_{1eff}+\tau_{IT}}{2}%
\end{array}
\right.  \;.
\end{equation}
We stress again that it is clearly seen that jumping may slow the search
considerably. Note that again the optimal value of $\tau_{1}$ is very
different than the canonical one discussed in Sec. \ref{Sliding and jumping}%
.\begin{figure}[ptb]
\begin{center}
\includegraphics[
height=3.0926in, width=4.7184in ]{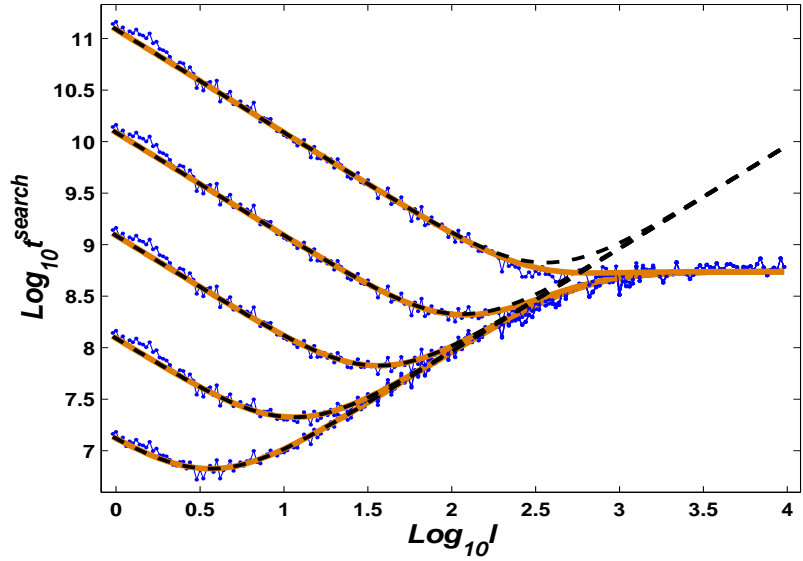}
\end{center}
\caption{The influence of ITs on the search time is shown. The search time,
$t^{search}$, is plotted vs. the antenna length, $l$, for a different values
of $\tau_{3}$ ($10,\ 100,\ 1000,\ 10000,\ 100000$ in units of $\delta$ from
bottom up). Here ITs, jumping and sliding are allowed. Thin solid lines with
dots represent the numerical results. The bold solid lines represent analytic
results (Eq. (\ref{tblRs})). The black, dashed lines represent the search time
in the case with no ITs, obtained by using Eq. (\ref{ts1}) with $D_{1eff}%
=\frac{R^{2}}{2\delta}$. Here $L$, $R$ and $b$ were taken to be $1224000$, $1$
and $20$ lattice constants respectively.}%
\label{fig14}%
\end{figure}Fig. \ref{fig14} shows a comparison between the theoretically
predicted search time (Eq. (\ref{tblRs})) and numerical simulation.

\section{Application to the Lac
repressor\label{Application to the Lac repressor}}

The above results cover a very wide variety of regimes. For a given protein
only several are of interest. To illustrate the use of the results presented
above we consider Lac repressor. Lac repressor is both the most studied
DNA-binding protein (see \cite{M96} for a review) and its structure is highly
suggestive of intersegmental transfers taking place. Despite of this several
physical parameters of the protein are yet unknown. In this subsection we use
the known parameters: $R\sim10nm$ \cite{RR97}, $\Lambda\sim1\mu$, $L\sim1mm$,
and those measured for Lac repressor with only one DNA-binding domain
$\tau_{1}\sim1ms$, $\tau_{3}\sim0.1\tau_{1}$ and $D_{1}\sim0.05\mu^{2}/s$
\cite{WAC2006,ELX2007}. Still unknown are $b$, the sliding length, and
$\tau_{IT}$ which we use as free parameters and study the search time as these
are varied. It is interesting to note that Lac repressor is so large that, as
we show, essentially all ITs can move the protein at each step to a completely
uncorrelated location on the DNA.

\begin{figure}[ptb]
\begin{center}
\includegraphics[
height=3.2387in, width=7.0041in ]{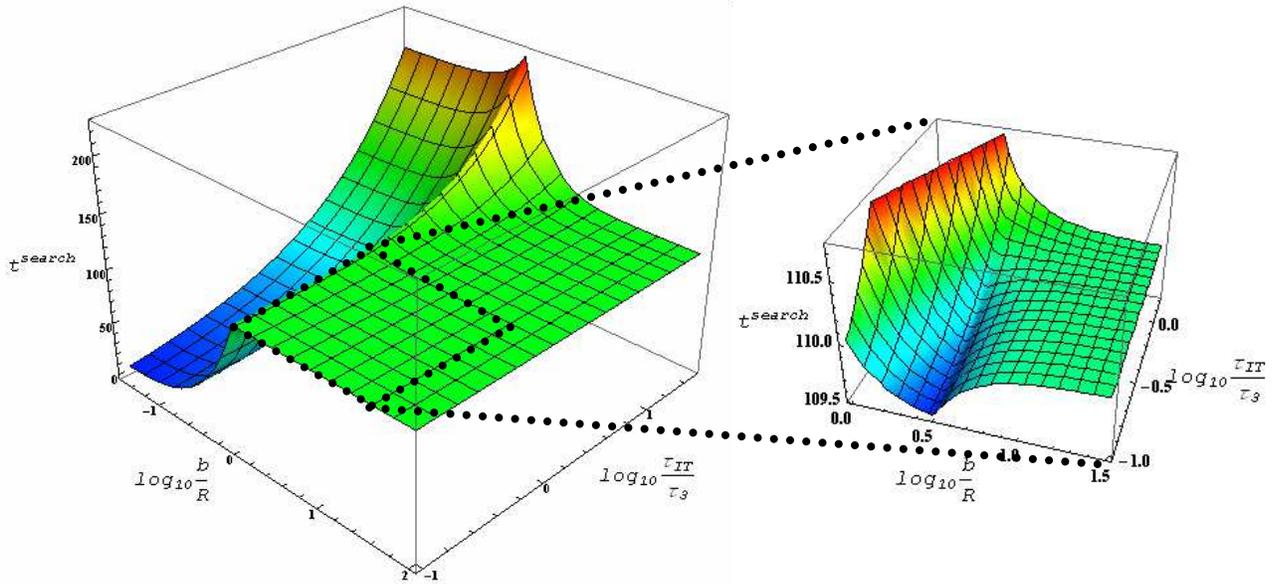}
\end{center}
\caption{On this figure the analytical prediction of $t^{search}$ is shown as
a function of the unknown parameters $b$ and $\tau_{IT}$.}%
\label{fig12}%
\end{figure}

Fig. \ref{fig12} shows the predicted $t^{search}$ from Secs.
\ref{Intersegmental transfer and jumping} and
\ref{Intersegmental transfer, sliding and jumping} as a function of $b$ and
$\tau_{IT}$. One may see that for $b\gg R$, ITs do not affect the search time
significantly even if $\tau_{IT}$ is small. This is results from the small
probability of performing UIT for a large values of $b$. However, if $b\ll R$
the search time may be decreased in a significant manner by including ITs. For
example, by setting $b$ to be the size of one base pair $\sim0.3nm$ the search
time decrease by a factor of three when $\tau_{IT}=\tau_{3}$ and if $\tau
_{IT}=\frac{\tau_{3}}{10}$ the search time decreases by a factor of ten.
Finally, Fig. \ref{fig12} shows that for large values of $\tau_{IT}$, ITs may
slow down the search process.

\section{Summary\label{Summary}}

In this article we presented a comprehensive study of the influence of ITs on
the search process. Using simple scaling arguments we studied a model which
includes the protein dynamics and DNA conformation. Two extreme regimes for
the DNA dynamics were studied: completely quenched (frozen) and annealed
(rapidly moving) DNA. ITs were assumed to relocate the protein to a randomly
chosen DNA position within a range of the order of the protein size. The
essence of the description may be understood from Sec.
\ref{Pure intersegmental transfer}. The following sections elaborate and study
a search processes based on ITs with sliding and/or jumping. The results for a
particular protein of interest may be obtained by suitably selecting the
section most relevant for a particular case.

The obtained results clearly indicate that including IT in the search process
may increase, the robustness of the search efficiency to different parameters
of the model such as the protein-DNA affinity, the three-dimensional diffusion
coefficient etc.

The mechanism of IT\ may produce a significant increase of the optimal
residence time of the protein on the DNA between two subsequent rounds of
three-dimensional diffusion from the value predicted by the models that do not
include IT. Recent experiments indicates that the value of the residence time
of the proteins on the DNA between two subsequent rounds of the
three-dimensional diffusion is much larger than the optimum predicted by the
model. It is possible that the existence of the IT mechanism may explain the
rather quick search times found \textit{in vivo} experiments.

One of the most surprising results found that above some critical value of the
typical time of a jump the protein has no reason to detach from the DNA. It is
more efficient for it to stay bound to the DNA. The value of the critical jump
time depends on the time of an IT.

A key ingredient needed for the behavior to occur is the confinement
of the DNA in a volume much smaller than its radius of gyration. The
probability to perform an UIT obviously depends on the DNA density.
Larger density implies a larger probability for UITs. Therefore the
effects of IT are expected to be more important in the systems with
high DNA density as cells or eucaryotic nuclei rather than in the
\textit{in vitro experiments. }

The dependency, mentioned above, on the DNA density leads to many
possible regimes which depend on the cell size, DNA length etc. In
particular, we found non-trivial regimes when the search time
increases as a square root of the DNA length or is completely
independent of it. Our estimates indicate that these seem to be the
ones most relevant to experiments.

Our results also show that the search on quenched and annealed DNA may have
quite different scaling behavior. In general a search that uses ITs is shown
to be more rapid on an annealed DNA than on a quenched DNA. This happens due
to the rapid decrease in correlations which results from the motion of the DNA molecule.

Similar scaling arguments were used to discuss the effects of IT in
\cite{HS2007}. However, there the main mechanism that drives the IT
was assumed to be the motion of the DNA molecule. In our study even
on completely quenched DNA ITs are shown to be important.

\begin{acknowledgments}
We thank R. Voituriez and R. Metzler for discussions and D. Levine
for discussions and comments on the manuscript. The Israel Science
Foundation is acknowledged for financial support.
\end{acknowledgments}

\begin{appendix}
\section{\label{Appendix A}}
In this appendix we argue that the typical time that the protein
spends in a jump is given by $\tau _{3}\sim \frac{\Lambda
^{3}}{D_{3}L}$. This quantity is controlled by average volume which
is free from DNA. Consider, first, the probability to find a volume,
free from DNA of radius $s$. To do so we describe the packed DNA as
an ideal gas of $\frac{L}{L_{0}}$\ straight rods
of length $L_{0}$ that are distributed randomly in the cell (see Fig. \ref%
{fig2}). The probability $p_{seg}$, that a given rods crosses a
volume of radius
$s$ is of order of $\frac{L_{0}^{3}}{\Lambda ^{3}}\frac{s^{2}}{L_{0}^{2}}=%
\frac{L_{0}s^{2}}{\Lambda ^{3}}$. Here $\frac{L_{0}^{3}}{\Lambda
^{3}}$ is the probability that a given segments is located within a
distance $L_{0}$ of a point inside the cell and $\sim
\frac{s^{2}}{L_{0}^{2}}$ is the probability that this segment
crosses a sphere of radius $s$ around the point. The probability
that at least one segment crosses the void is
\begin{equation}
1-\left( 1-p_{seg}\right) ^{L/L_{0}}\simeq 1-e^{-\frac{LR^{2}}{\Lambda ^{3}}%
}\;.
\end{equation}
Therefore the typical free volume radius is $\sim
\sqrt{\frac{\Lambda ^{3}}{L}} $. Hence, the typical time to
explore\footnote{In the three-dimensional space diffusive
exploration is not compact i.e. the probability to find a finite
target (sphere) is less than one. However, the DNA as a target may
be described as a set of straight rods. Hence, the search process
effectively looks like the two-dimensional search for a finite
target (disk) i.e. compact (up to logarithmic corrections).} this
volume is $\tau _{3}\sim \frac{\Lambda ^{3}}{D_{3}L}$. A second way
to get the same expression for $\tau_3$ is based on a comparison
between Eqs. (\ref{t3D}) and (\ref{ts1}). Obviously, in the limiting
case $\tau _{1}\ll \tau _{3}$ and $\sqrt{2D_{1}\tau _{1}}=r$, the
search becomes based only on the three-dimensional diffusion. Hence,
in this case the formula (\ref{ts1}) should give (\ref{t3D}). It is
easy to see that this happens only when $\tau _{3}\sim \frac{\Lambda
^{3}}{D_{3}L}$.
\section{\label{Appendix B}}
In this appendix we describe the details of the numerical
simulation. The simulations were done on a cubic lattice containing
$800\times 800\times 800$ sites. Assuming that a real cell has a
volume of $1\mu m^{3}$ each site on
the lattice represents a volume of $\left( dx\right) ^{3}=\left( \frac{\mu m%
}{800}\right) ^{3}$. Polymers (representing the DNA) with different
lengths were embedded in the lattice by using a self-avoiding random
walk. The persistence length was accounted for by assigning a
probability $p_{0}$ of changing direction randomly among the
possible directions. Using the persistence length of about $50nm$
leads to $p_{0}=\frac{dx}{50nm}=0.025$. If during the process of
generating the configuration the polymer length can not be extended
we shrink the polymer by $O\left( 10\right) $ lattice constants and
regenerate. While this leads to a bias in the configuration for
single realization confined in a box, which are of interest, we
expect no effect on the results (a non biased algorithm is not
plausible within our computational resources). The search process is
simulated following the model described in the text. In each step
the protein has a probability $\frac{dx^{2}}{2b^{2}}$ to perform an
IT and a probability $\frac{dx^{2}}{2l^{2}}$ to perform a jump. ITs
were simulated by a randomly choosing a DNA site within a distance
$R$ from the location of the protein. With the exception of Sec.
\ref{Sliding and jumping}, where a complete simulation of the
three-dimensional diffusion was carried out by performing moves to
the $6$ available directions, a jump was simulated by randomly
choosing a site on the DNA. The time of the jump was taken as a free
constant ().
\section{\label{Appendix C}}
In this appendix we argue that using ITs the protein can only move
along the
chemical coordinate to distances smaller than $R$ or larger than $\frac{%
\Lambda ^{2}}{L_{0}}$. As mentioned above, we assume that during an
IT the protein chooses a new location whose three-dimensional
distance from its current location is smaller than $R$. The new
location is chosen randomly with a uniform probability. Given the
uniform probability we need to estimate the total typical length
available at each IT, $G$. We separate \ this quantity to four types
of contributions:
\begin{equation}
G=G_{1}+G_{2}+G_{3}+G_{4}\;.
\end{equation}%
The first $G_{1}$ is the contribution from DNA whose distance along
the chemical coordinate from a point $x$ is smaller than $R$, the
protein size. This is given by
\begin{equation}
G_{1}\left( x\right) \simeq 2R\;.  \label{G1}
\end{equation}
The contribution $G_{2}$ arises from DNA whose chemical distance
from a point $x$ is larger than $R$ but smaller than $L_{0}$. The
probability for
the DNA to bend on a scale $l$ is approximately given by $\frac{%
1-e^{-l/2L_{0}}}{L_{0}}$. However, the probability that this\ bend
will connect to $x$ is $\sim \frac{R^{2}}{l^{2}}$ (due to the area
ratio). Since each connection contributes a length of the order of$\
R$ to $G_{2}$ we obtain
\begin{equation}
G_{2}\sim R{\int\limits_{R}^{L_{0}}}\frac{1-e^{-l/2L_{0}}}{L_{0}}\frac{R^{2}%
}{l^{2}}dl\sim \frac{R^{2}}{L_{0}}\left( 1-e^{-R/2L_{0}}\right) \simeq \frac{%
R^{3}}{L_{0}^{2}}\;.
\end{equation}
The contribution $G_{3}$ comes from DNA whose chemical distance from
$x$ is larger than $L_{0}$ but smaller than the length at which the
DNA feels the boundaries of the cell is  $\sim \frac{\Lambda
^{2}}{L_{0}}$. This value can be overestimated using the fact that a
free three-dimensional random walk \textit{on a lattice} returns to
the origin about $1.5$ times on average. Therefore, a continuous
free three-dimensional random walk with persistence
length $L_{0}$ returns to a region with radius $R$ an order of $\left( \frac{%
R}{L_{0}}\right) ^{2}$ times. Each such return contributes length of
about $R$ to $G_{3}$, leading to
\begin{equation}
G_{3}\sim \frac{R^{3}}{L_{0}^{2}}\;.  \label{G3}
\end{equation}
Finally, $G_{4}$ is the contribution from the rest of the DNA (whose
chemical distance is larger than $\frac{\Lambda ^{2}}{L_{0}}$ but
smaller than $L$). Using (\ref{pseg}) and since each connected
segment contributes a length of the order $R$ to $G_{4}$ one obtains
\begin{equation}
G_{4}\sim R\frac{L}{L_{0}}p_{seg}\sim \frac{LR^{3}}{\Lambda ^{3}}\;.
\label{G4}
\end{equation}%
This result can be understood within a mean field approach: if the
DNA has a total length $L$ and is assumed to be distributed
uniformly in the cell, every volume in the cell contains a part of
the total DNA length that is equal to the total DNA length times the
fraction of the volume. One can see that in the assumed regime where
$L_{0}\gg R$ and $L\gg \Lambda
\frac{\Lambda ^{2}}{L_{0}^{2}}$, $G_{2}$ and $G_{3}$ are much smaller than $%
G_{4}$ and $G_{1}$. Therefore, we can safely neglect the probability
that the protein will move to a location on the DNA whose chemical
distance from protein's actual location is larger than $R$ and
smaller than $\frac{\Lambda ^{2}}{L_{0}}$.
\section{\label{D}}
In this appendix\ the effective times $\tau _{1eff}$ and $\tau _{3eff}$ are
calculated.
\subsection{The effective time of a correlated movement\label{D1}}
We have two independent mechanisms for an uncorrelated motion. The
first is jumping with a typical time of $\tau _{1}$ between two
subsequent jumps. This process has Poissonian statistics and
therefore the probability that the protein does not perform a jump
before time $t$ is
\begin{equation}
P_{J}=\exp \left( -\frac{t}{\tau _{1}}\right)\; .
\end{equation}%
The second mechanism for uncorrelated motion is an UIT with a
typical time of order of $\frac{\lambda ^{2}}{D_{1eff}}$ between two
subsequent UITs. In the case of annealed DNA this mechanism has
Poissonian statistics and the probability that the protein does not
perform an UIT before time $t$ is
\begin{equation}
P_{IT}\sim \exp \left( -\frac{t}{\lambda ^{2}/D_{1eff}}\right)\;.
\end{equation}%
For quenched DNA the probability that the protein did not performed
an UIT after traveling distance $x$ is $\sim e^{-x/\lambda }$. Since
the protein performs an effective one-dimensional diffusion, $x\sim
\sqrt{2D_{1eff}t}$ and we obtain
\begin{equation*}
P_{IT}\sim \exp \left( -\sqrt{\frac{t}{\lambda
^{2}/2D_{1eff}}}\right)\; .
\end{equation*}%
We will take the typical time of a non-interrupted (by an uncorrelated
relocation) one-dimensional effective diffusion to be
\begin{equation}
\tau _{1eff}=\int_{0}^{\infty }P_{IT}P_{J}dt\sim \frac{1}{2}\frac{1}{\frac{%
D_{1eff}}{l^{2}}+\frac{D_{1eff}}{\lambda ^{2}}}\;.
\end{equation}%
The last expression is exact in the annealed case but it is only an
approximation in the quenched regime. One can verify that the error does not
exceed $50\%$, which is sufficient for scaling arguments of the type used in
the paper.
\subsection{The effective time of an uncorrelated movement\label{D2}}
Since there are two mechanisms for uncorrelated movement: a jump
with a typical time $\tau_3$ and an UIT with a typical time $\delta$
the typical time of the uncorrelated movement is the average of
$\tau_3$ and $\delta$ weighted by the relevant probabilities for
each process:
\begin{align}
\tau _{3eff}& =\delta \int_{0}^{\infty }\left( -\frac{dP_{IT}}{dt}\right)
P_{J}dt+\tau _{3}\int_{0}^{\infty }\left( -\frac{dP_{J}}{dt}\right) P_{IT}dt=
\notag \\
& =\delta \int_{0}^{\infty }\frac{dP_{J}}{dt}P_{IT}dt-\tau
_{3}\int_{0}^{\infty }\frac{dP_{J}}{dt}P_{IT}dt-\delta \int_{0}^{\infty }%
\frac{d}{dt}\left( P_{J}P_{IT}\right) dt=  \notag \\
& =\frac{\tau _{3}-\delta }{\tau _{1}}\int_{0}^{\infty }P_{J}P_{IT}dt+\delta
=\frac{\tau _{3}-\delta }{\tau _{1}}\tau _{1eff}+\delta =  \notag \\
& =\tau _{3}\frac{\tau _{1eff}}{\tau _{1}}+\delta \left( 1-\frac{\tau _{1eff}%
}{\tau _{1}}\right) =\tau _{3}\frac{l_{eff}^{2}}{l^{2}}+\delta \left( 1-%
\frac{l_{eff}^{2}}{l^{2}}\right) =\frac{\tau _{3}+\delta \frac{l^{2}}{%
\lambda ^{2}}}{1+\frac{l^{2}}{\lambda ^{2}}}\;.
\end{align}%
In the case of sliding $\delta $ is replaced by $\tau _{IT}$.%
\end{appendix}

\bibliographystyle{unsrt}
\bibliography{misha1,misha2}

\begin{thebibliography}{10}

\bibitem{ABLRRW94}
B.~Alberts, D.~Bray, J.~Lewis, M.~Raff, K.~Roberts, and J.~D. Watson.
\newblock {\em The molecular biology of the cell}.
\newblock Garland, New York, 4'th edition, 1994.

\bibitem{LBMV2008}
C.~Loverdo, O.~Benichou, M.~Moreau, and R.~Voituriez.
\newblock Enhanced reaction kinetics in biological cells.
\newblock {\em Nature Physics}, 4:134, 2007.

\bibitem{S17}
M.~von Smoluchowski.
\newblock Mathematical theory of the kinetics of the coagulation of colloidal
  solutions.
\newblock {\em Z. Phys. Chem.}, 92:129, 1917.

\bibitem{ESWSL99}
M.~B. Elowitz, M.~G. Surette, P.~E. Wolf, J.~B. Stock, and S.~Leibler.
\newblock Protein mobility in the cytoplasm of {E}scherichia coli.
\newblock {\em J. Bacteriol.}, 181(1):197, 1999.

\bibitem{LR72}
S.~Y. Lin and A.~D. Riggs.
\newblock {L}ac repressor binding to non-operator {DNA}: detailed studies and a
  comparison of eequilibrium and rate competition methods.
\newblock {\em J. Mol. Biol.}, 72(3):671, 1972.

\bibitem{RSB70}
A.~D. Riggs, H.~Suzuki, and S.~Bourgeois.
\newblock {L}ac repressor-operator interaction {I}. {E}quilibrium studies.
\newblock {\em J. Mol. Biol.}, 48(1):67, 1970.

\bibitem{RBC70}
A.~D. Riggs, S.~Bourgeois, and M.~Cohn.
\newblock The {L}ac repressor-operator interaction. 3. {K}inetic studies.
\newblock {\em J. Mol. Biol.}, 53(3):401, 1970.

\bibitem{BWH81}
O.~G. Berg, R.~B. Winter, and P.~H. von Hippel.
\newblock Diffusion-driven mechanisms of protein translocation on nucleic
  acids. 1. models and theory.
\newblock {\em Biochemistry}, 20(24):6929, 1981.

\bibitem{SM2004}
M.~Slutsky and L.~A. Mirny.
\newblock Kinetics of protein-{DNA} interaction: Facilitated target location in
  sequence-dependent potential.
\newblock {\em Biophys J.}, 87:4021, 2004.

\bibitem{HGS2006}
T.~Hu, A.~Y. Grosberg, and B.~I. Shklovskii.
\newblock How proteins search for their specific sites on {DNA}: The role of
  {DNA} conformation.
\newblock {\em Biophys J.}, 90:2731, 2006.

\bibitem{HS2007}
Tao Hu and B.~I. Shklovskii.
\newblock How proteins search for their specific sites on {DNA}: The role of
  intersegment transfer.
\newblock {\em Phys. Rev. E}, 76:051909, 2007.

\bibitem{BB76}
O.~G. Berg and C.~Blomberg.
\newblock Association kinetics with coupled diffusional flows. special
  application to the {L}ac repressor-operator system.
\newblock {\em Biophys. Chem.}, 4:367, 1976.

\bibitem{HM2004}
S.~E. Halford and J.~F. Marko.
\newblock How do site-specific {DNA}-binding proteins find their targets?
\newblock {\em Nucleic Acids Research}, 32(10):3040, 2004.

\bibitem{BZ2004}
B.P. Belotserkovskii and D.A. Zarling.
\newblock Analysis of a one-dimensional random walk with irreversible losses at
  each step: applications for protein movement on {DNA}.
\newblock {\em J. Theor. Biol.}, 226:195, 2004.

\bibitem{Z2005}
H.~X. Zhou.
\newblock A model for the mediation of processivity of {DNA}-targeting proteins
  by nonspecific binding: Dependence on {DNA} length and presence of obstacles.
\newblock {\em Biophysical Journal}, 88:1608, 2005.

\bibitem{LAM2005}
M.~A. Lomholt, T.~Ambjrnsson, and R.~Metzler.
\newblock Optimal target search on a fast-folding polymer chain with volume
  exchange.
\newblock {\em Phys. Rev. Lett.}, 95:260603, 2005.

\bibitem{AD68}
G.~Adam and M.~Delbruck.
\newblock Reduction of dimensionality in biological diffusion processes.
\newblock In A.~Rich and N.~Davidson, editors, {\em Structural Chemistry and
  Molecular Biology}, pages 198--215, San Francisco, CA, 1968. Freeman.

\bibitem{H95Volume1}
B.~D. Hughes.
\newblock {\em Random walks and random enviroments}, volume 1: Random walks.
\newblock Clarendon press, Oxford, UK, 1995.

\bibitem{WAC2006}
Y.~M. Wang, Robert~H. Austin, and Edvard~C. Cox.
\newblock Single molecule measurement of repressor protein 1d diffusion on
  {DNA}.
\newblock {\em Phys.Rev. Lett.}, 97:048302, 2006.

\bibitem{ELX2007}
J.~Elf, G.-W. Li, and P.~X. Xie.
\newblock Probing transcription factor dynamics at the simple single-molecule
  level in a living cell.
\newblock {\em Science}, 316:1191, 2007.

\bibitem{K-HRBOCNVH77}
Y.~Kao-Huang, A.~Revzin, A.~P. Butler, P.~O'Conner, D.~W. Noble, and P.~H.~Von
  Hippel.
\newblock Nonspecific {DNA} binding of genome-regulating proteins as a
  biological control mechanism: Measurement of {DNA}-bound {E}scherichia coli
  {L}ac repressor in vivo.
\newblock {\em PNAS}, 74:4228, 1977.

\bibitem{HRGW75}
P.~H. von Hippel, A.~Revzin, C.~A. Gross, and A.~C. Wang.
\newblock In H.~Sund and G.~Blauer, editors, {\em Protein-Ligand Interactions},
  pages 279--347, Berlin, 1975. Walter de Gruyter.

\bibitem{BC75}
J.~L. Bresloff and D.~M. Crothers.
\newblock {DNA}-ethidium reaction kinetics: demonstration of direct ligand
  transfer between {DNA} binding sites.
\newblock {\em L. Mol. Biol.}, 172:263, 1975.

\bibitem{FM92}
R.~Fickert and B.~Muller-Hill.
\newblock How {L}ac repressor finds {L}ac operator in vitro.
\newblock {\em J. Mol. Biol.}, 226(1):59, 1992.

\bibitem{RC92}
T.~Ruusala and D.~M. Crothers.
\newblock Sliding and intermolecular transfer of the {L}ac repressor: Kinetic
  perturbation of a reaction intermediate by a distant {DNA} sequence.
\newblock {\em Proc. Natl. Acad. Sci. USA}, 89:4903, 1992.

\bibitem{LN97}
B.~A. Lieberman and S.~K. Nordeen.
\newblock {DNA} intersegment transfer, how steroid receptors search for a
  target site.
\newblock {\em J. Biol. Chem.}, 272(2):1061, 1997.

\bibitem{EWWH99}
M.~L. Embleton, S.~A. Williams, M.~A. Watson, and S.~E. Halford.
\newblock Specificity from the synapsis of {DNA} elements by the {S}fi{I}
  endonuclease.
\newblock {\em J. Mol. Biol.}, 289(4):785, 1999.

\bibitem{BGZY99}
C.~Bustamante, M.~Guthold, X.~Zhu, and G.~Yang.
\newblock Facilitated target location on {DNA} by individual {E}scherichia coli
  {RNA} polymerase molecules observed with the scanning force microscope
  operating in liquid.
\newblock {\em ASBMB}, 274(24):16665, 1999.

\bibitem{FC84}
M.~G. Fried and D.~M. Crothers.
\newblock Kinetics and mechanism in the reaction of gene regulatory proteins
  with {DNA}.
\newblock {\em J. Mol. Biol.}, 172:263, 1984.

\bibitem{CBTVK2007}
S.~Condamin, O.~Be´nichou, V.~Tejedor, R.~Voituriez, and J.~Klafter.
\newblock First-passage times in complex scale-invariant media.
\newblock {\em Nature}, 450:77, 2007.

\bibitem{LR75}
S.~Y. Lin and A.~D. Riggs.
\newblock The general affinity of {L}ac repressor for {E}. coli {DNA}:
  implication for gene regulation in procaryotes and eucaryotes.
\newblock {\em Cell}, 4:107, 1975.

\bibitem{RCMKAFR87}
B.~Richey, D.~S. Cayley, M.~C. Mossing, C.~Kolka, C.~F. Anderson, T.~C. Farrar,
  and M.~T. Record.
\newblock Variability of the intracellular ionic environment of {E}scherichia
  coli. differences between in vitro and in vivo effects of ion concentrations
  on protein-{DNA} interactions and gene expression.
\newblock {\em J. Biol. Chem}, 262:7157, 1987.

\bibitem{W99SM}
D.~J. Watts.
\newblock {\em Small Worlds: The Dynamics of Networks Between Order and
  Randomness.}
\newblock Princeton University Press, 1999.

\bibitem{M96}
M.~Muller-Hill.
\newblock {\em The {L}ac operon. A short history of a genetic paradigm}.
\newblock Walter de Gruyter, Berlin, 1996.

\bibitem{RR97}
G.~C. Ruben and T.~B. Roos.
\newblock Conformation of {L}ac repressor tetramer in solution, bound and
  unbound to operator {DNA}.
\newblock {\em Microscopy Research and Technique}, 36:400, 1997.

\end{thebibliography}

\end{document}